\documentclass[preprint,review,12pt]{elsarticle}

\usepackage[colorinlistoftodos]{todonotes}
\usepackage{placeins}

\usepackage{amssymb}
\usepackage{mhchem}
\usepackage{overpic}
\usepackage{graphics}
\usepackage{multirow}
\usepackage{rotating}
\usepackage{textcomp}
\usepackage{float}
\usepackage{color}
\usepackage{subfigure}
\usepackage{soul}

\journal{Desalination}

\begin{document}

\begin{frontmatter}
\title{Flow-electrode capacitive deionization enables continuous and energy-efficient brine concentration}

\author[DWI,AVT.CVT]{Alexandra Rommerskirchen}
\author[DWI,AVT.CVT]{Christian J. Linnartz}
\author[DWI,AVT.CVT]{Franziska Egidi}
\author[AVT.CVT]{Sefkan Kendir}
\author[DWI,AVT.CVT]{Matthias Wessling\corref{mycorrespondingauthor}}
\cortext[mycorrespondingauthor]{Phone: +49-241-8095470, Fax: +49-241-8092252, Manuscripts.cvt@avt.rwth-aachen.de, Forckenbeckstra\ss e 51, 52074 Aachen, Germany}

\address[DWI]{DWI - Leibniz-Institute for Interactive Materials e.V., Forckenbeckstra\ss e 50, 52074 Aachen, Germany}
\address[AVT.CVT]{RWTH Aachen University, Aachener Verfahrenstechnik-Chemical Process Engineering, Forckenbeckstra\ss e 51, 52074 Aachen, Germany}

\begin{abstract}
Many industrial and agricultural applications require the treatment of water streams containing high concentrations of ionic species for closing material cycles. High concentration factors are often desired, but hard to achieve with established thermal or membrane-based water treatment technologies at low energy consumptions. Capacitive deionization processes are normally assumed as relevant for the treatment of low salinity solutions only. Flow-electrode capacitive deionization (FCDI), on the other hand, is an electrically driven water desalination technology, which allows the continuous desalination and concentration of saline water streams even at elevated salinities. Ions are adsorbed electrostatically in pumpable carbon flow electrodes, which enable a range of new process designs. 

In this article, it is shown that continuously operated FCDI systems can be applied for the treatment of salt brines. Concentrations of up to 291.5~g/L NaCl were reached in the concentrate product stream. Based on this, FCDI is a promising technology for brine treatment and salt recovery. Additionally, a reduction of the energy demand by more than 70\% is demonstrated by introducing  multiple cell pairs into a continuous FCDI system. While the economic feasibility is not investigated here, the results show that FCDI systems may compete with established technologies regarding their energy demand.
\end{abstract}

\end{frontmatter}


\section{Introduction and motivation} 

Water treatment technologies should be energy efficient, reliable and economical. While the technologies available for water treatment, such as desalination and concentration have been improved significantly over the last decades, there is no viable water treatment technology available in some application cases, while in many other cases there is still room for improvement \cite{Elimelech2011,Anderson2010,Porada2013Review}. \\

Significant attention has been focused on capacitive deionization (CDI) technologies in the last decade. In CDI processes, an applied electrochemical potential gradient leads to the transport of dissolved ions towards the oppositely charged electrodes, where the ions are bound via electrostatic adsorption at the electrode-solution interface. High surface areas, and hence high ion adsorption capacities, are usually achieved by employing solid, porous carbon electrodes. CDI processes can be operated in a variety of configurations. The use of ion-exchange membranes (IEM) can improve the performance of CDI processes regarding their desalination performance and energy efficiency, when placed in between electrodes and treated solution \cite{Anderson2010,Porada2013Review,Oren2008,Suss2015Review,AlMarzooqi2014,Tang2019a}.
Most CDI processes are based on static carbon-based electrodes and are usually operated discontinuously using charge-discharge cycles. Due to this, the application of CDI processes is usually limited to the treatment of brackish water feed streams. Higher feed salinities would lead to a fast saturation of the electrode material with ions and hence the requirement of very short charge-discharge cycles, which would make CDI processes with static electrodes inefficient due to back-mixing withing the tubing and modules.\\ 

An upcoming CDI technology suitable for the treatment of higher salinities is flow-electrode capacitive deionization (FCDI) \cite{Jeon2013Desalination}. In case of FCDI, flow electrodes consisting of pumpable carbon suspensions replace the static CDI electrodes, in which charge percolation leads to a utilization of the suspended carbon particles as electrode material \cite{Akuzum2020,Shukla2019}. Flow electrodes can be regenerated via mixing \cite{Jeon2014Ion} or in a separate module or compartment, and enable a fully continuous operation. By now, various continuous FCDI process layouts have been proposed, including a continuous two-module configuration \cite{Gendel2014}, single-module configuration \cite{Rommerskirchen2015}, a mixer-settler configuration \cite{Doornbusch2016}, and a two-step regeneration configuration \cite{Rommerskirchen2018Energy}. In all of these configurations, the electrodes are continuously regenerated and a diluate and concentrate stream is produced continuously with more or less constant salt concentrations. 

FCDI technologies are a promising research field due to the high salt adsorption capacities \cite{Suss2015Review,Gendel2014,Shanbhag2016,Hatzell2015Materials}, good desalination performance, the ability to desalinate high salinity solutions \cite{Jeon2013Desalination}, and the possibility of fully continuous operation \cite{Gendel2014,Rommerskirchen2015,Doornbusch2016,Dahiya2020}. A great strength of the FCDI technology is its versatility. The use of suspension-based flow electrodes for desalination and water treatment in general can be seen as a platform technology, which can be operated in many different process layouts for many different applications \cite{Porada2014Carbon,Hatzell2014FCDIconcept}. Lab-scale FCDI studies published until now were, for example, aimed at salt metathesis \cite{Linnartz2017}, ammonia recovery from wastewaters \cite{Zhang2018FCDIStripping,Zhang2019Ammonia,Fang2018}, nitrogen and phosphorous recovery \cite{Bian2019}, nitrate removal \cite{Song2019}, copper removal \cite{Zhang2019CuFCDI}, and uranium removal \cite{Ma2019}. Apart from a multitude of pure FCDI processes, many combinations with established processes are imaginable, such as the combination of FCDI with nanofiltration \cite{Choi2017hybridNF}. However, most of the prior work on FCDI focuses on low or medium ionic concentration applications, for brackish water and seawater treatment, due to the limited adsorption capacity of the carbon electrodes in discontinuous or semi-continuous operation of FCDI processes.\\

Several characteristics make FCDI a promising candidate for the treatment of high salinity solutions. Compared to most established thermal or membrane-based water treatment processes, FCDI processes may enable the selective removal of charged species from water streams in the future, and may at the same time become more energy-efficient for many application cases. Even in comparison to other electrically driven processes \cite{Giwa2017}, FCDI has its advantages: Unlike electrodialysis (ED) processes, FCDI relies on the capacitive adsorption of ions in the electrodes at low voltages, and does not require electrochemical reactions in the electrodes. Hence, gas evolution and the use of acids in the electrode compartments is avoided. The elimination of electrochemical reactions in the electrode compartments may also be advantageous for the treatment of sensitive chemicals, which may be altered by the high potentials and side products in case of ED processes. 

As shown in our previous article \cite{Rommerskirchen2018Energy}, the values regarding the energy demand of FCDI processes cannot yet compete with literature values regarding existing technologies. 
However, a target-oriented advancement of the FCDI technology may lead to a new alternative solution for many water treatment applications.

This article demonstrates the possibility of concentrating brines in a continuous process based on a single module using FCDI technology. Additionally, it is shown that a significant reduction of the energy demand and an improved productivity can be achieved by using several cell pairs. This approach is similar to the approach recently presented by Ma et al. \cite{Ma2020}, who presented an FCDI configuration with stacked membranes, which leads to significant productivity improvements, but only enables a semi-continuous water treatment.
In our work, we demonstrate the continuous treatment of saline brine using FCDI modules with one or more cell pairs, and therefore an even number of water compartments. 

Apart from the investigation of the electrical energy demand for desalination, we also included a comparison to the contribution of the pumping energy to the total energy demand in Section~\ref{comparison_desalination_pumpingenergy}.
Finally, we compare the presented FCDI concept to the established technologies mechanical vapor compression (MVC) and electrodialysis (ED) regarding their energy demand, which presents a complement to the study recently published by Patel et al. \cite{Patel2020}.
While the experiments presented are specifically aimed at the treatment and concentration of high salinity solutions (brines) using FCDI processes, the findings can be transferred to other potential application areas of FCDI processes.


\subsection{Reducing the resistance losses in FCDI processes: An analysis}
For a better understanding of the contribution of individual electrical resistances in an FCDI cell, theoretical estimations of the individual resistances were made. 

\begin{figure}[t]
	\centering
    \subfigure{\includegraphics[width=0.49\textwidth]{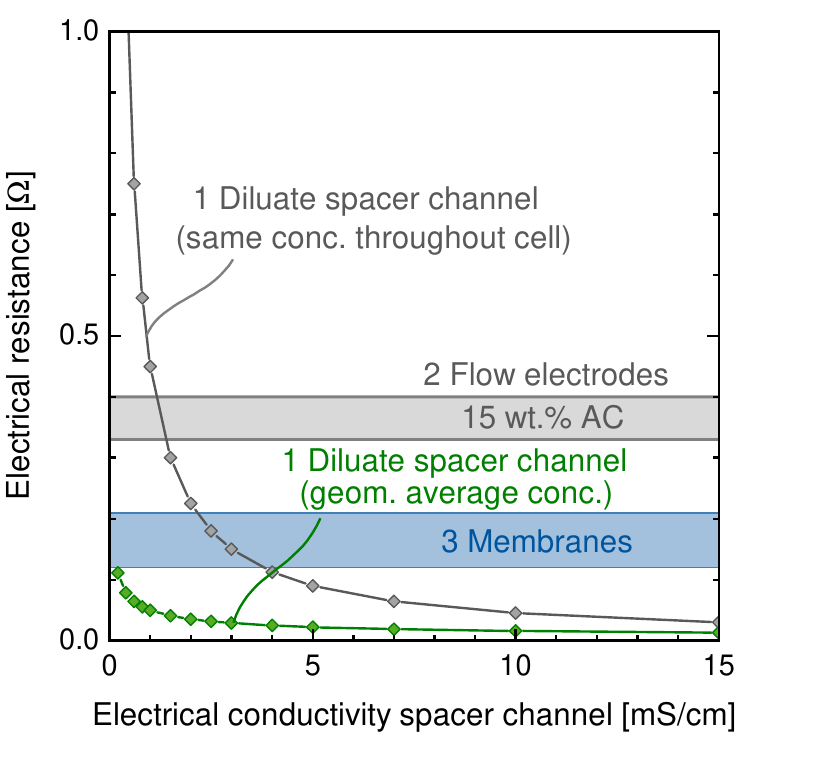}}
\caption{Results of theoretical calculations regarding the influence of the NaCl concentration/ electrical conductivity of the diluate stream on the electrical resistances in an FCDI cell with an active area of 100~cm$^2$.  Assumptions: Constant diluate flow rate of 1~mL/min, diluate concentration constant throughout cell (overestimated resistances/cell voltage), no consideration of laminar boundary layers, desalination up to the required concentration (varied electrical current).}
\label{fig:FCDI_influences_1}
\end{figure}

The results are plotted in Figure~\ref{fig:FCDI_influences_1}, which shows the electrical resistance of specific parts of an FCDI cell depending on the electrical conductivity of the diluate product stream. The calculations are based on the following assumptions:
\begin{itemize} 
\setlength\itemsep{0.001em}
\item NaCl feed: 60~g/L NaCl  at a diluate feed flow rate of 1~mL/min.
\item Resistance of membranes and flow electrodes constant throughout cell.
\item The required electrical current is calculated based on the desalination down to the respective diluate concentration, leading to lower energy demands at higher diluate concentrations even at constant resistance.
\item (Depleted) laminar boundary layers are not considered.
\item Ideal pumping energy: Pump efficiency is not considered.
\item Diluate channel resistance: Geometric average between inlet and outlet concentrations.
\item A current efficiency of one is assumed.
\item Flow-electrode resistance $R_{Fe}$: based on total resistance $R_{total}$ and combined solution/ membrane resistance R$_{S}$ from EIS measurements presented in \cite{Rommerskirchen2019}, $R_{Fe}=R_{total}-R_{S}$. 
\end{itemize}

The resistance of the flow electrodes is 0.33-0.4~$\Omega$ for an AC content of 15~wt.\%, which presents a large fraction of the overall cell resistance. This is in agreement with our previous study \cite{Rommerskirchen2019}, in which we observed a significant influence of the carbon slurry quality on the desalination performance.

The second largest resistance in the cell at most diluate concentrations is presented by the membranes. Three ion-exchange membranes of the type fumatech FAB/FKB with PEEK reinforcements, which we currently use in FCDI systems for the treatment of high salinity solutions, account for around 0.12-0.21~$\Omega$ of the total cell resistance, based on the manufacturers data. 
The resistance of the diluate channel varies significantly depending on the outlet concentration, ranging between 0.02-0.1~$\Omega$. The sum of the membrane resistance and the resistance of the diluate spacer channel should be equal to the resistance R$_{S}$ from the equivalent circuit model presented in our previous article \cite{Rommerskirchen2019}. In this, R$_{S}$ was usually found to be in the range of 0.2-0.27~$\Omega$, which is in good agreement with the estimations presented here. In case of most diluate concentrations, the membrane resistance dominates over the solution resistance, especially when considering three membranes in an FCDI single module configuration with one cell pair. At very low concentrations, the solution resistance is in a similar range as the membrane resistance. However, the solution resistance in this case is likely underestimated, as the laminar boundary layers are not considered.\\

Based on this analysis, the reduction of resistances in FCDI cells should be prioritized in the following order: (1) flow electrodes, (2) membranes and (3) diluate channel. Depending on the aimed-at diluate product concentration and the depletion of the diluate boundary layers, the reduction of the diluate channel resistance should be prioritized over the membrane resistance in specific cases.\\

Many studies have been published in CDI and FCDI literature, which focus on the selection of the active material and the overall flow-electrode composition to achieve a reduction of the flow-electrode resistance \cite{Porada2014Carbon,Hatzell2015Effect,Liang2017,Yang2017Plate,Tang2019,Cho2019CNT}. Here, we present a different approach aiming at the module design, based on the use of multiple cell pairs. The hypothesis is that the number of desalination units per unit flow electrode should be increased in order to reduce the relative contribution of the flow electrodes to the total resistance.

\subsection{Concept: FCDI with multiple cell pairs} \label{Concept}

\begin{figure}[t]
	\centering
    \subfigure{\includegraphics[width=0.45\textwidth]{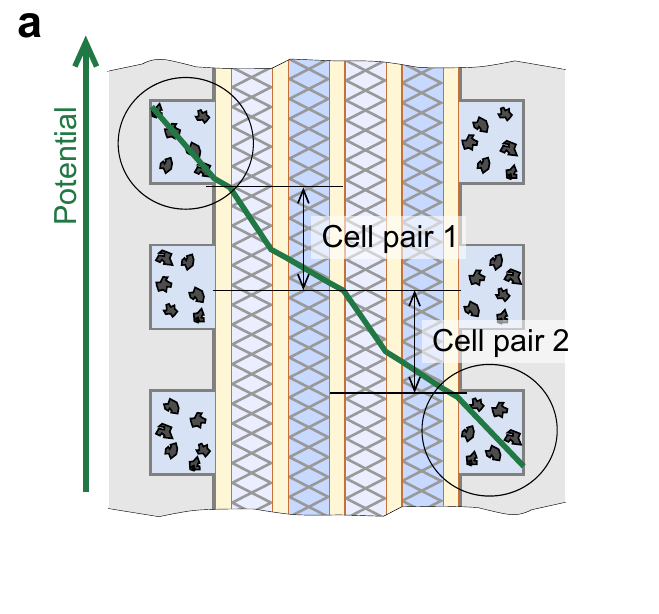}}
    \subfigure{\includegraphics[width=0.45\textwidth]{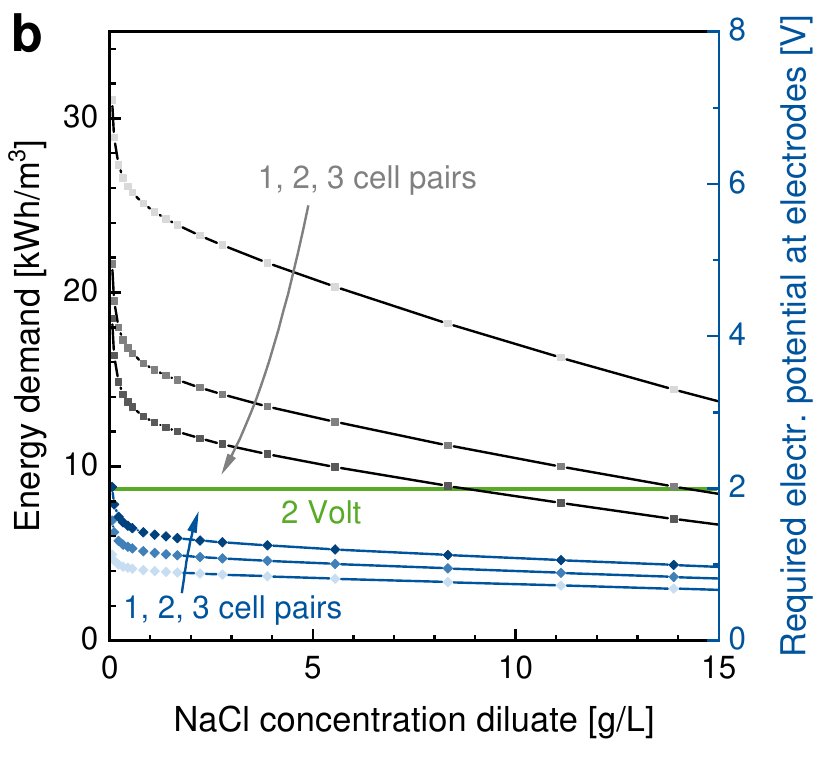}}
\caption{Concept of using multiple cell pairs for reducing the energy demand of continuous FCDI processes. (a) Illustration of the concept. The division of the flow-electrode voltage drop by the number of cell pairs leads to a reduced average voltage drop per cell pair. (b) Results of theoretical calculations regarding the influence of the NaCl concentration/electrical conductivity of the diluate stream on the energy demand and required cell potential in an FCDI cell with 1, 2 and 3 cell pairs.  Assumptions: Constant diluate flow rate of 1~mL/min, diluate concentration constant throughout cell (overestimated resistances/cell voltage), no consideration of laminar boundary layers, desalination up to the required concentration (varied electrical current). }
\label{fig:concept_stack}
\end{figure}

The motivation for introducing additional cell pairs into an FCDI module is illustrated in Figure~\ref{fig:concept_stack}~(a). By introducing one or more additional cell pairs (each one diluate and one concentrate channel, two membranes), the influence of the voltage drop over the electrodes is reduced. The flow rate of the feed stream is multiplied by the number of cell pairs, to achieve similar flow conditions within the FCDI module. While the overall ohmic resistance of the FCDI module increases with increasing number of cell pairs, there is only one pair of flow electrodes required. Hence, the ohmic resistance of the flow electrodes is divided by the number of cell pairs, and the overall impact of the electrode resistance is reduced. The theoretical estimations plotted in Figure~\ref{fig:concept_stack}~(b) show results regarding the energy demand and the expected required cell voltage for the desalination down to the desired diluate concentration (x-axis) in an FCDI system with one, two and three cell pairs. The calculations are based on the same assumptions as described above.

The calculations predict a steep decrease in the energy demand with increasing number of cell pairs, the most significant decrease being expected when introducing the second cell pair. Limitation for the number of cell pairs is the voltage applied to the graphite current collectors. At high voltages, unwanted faradaic reactions (e.g. carbon oxidation) may occur. In this work, a maximum of 2~V was applied in longer lasting experiments, indicated by the green line in Figure~\ref{fig:concept_stack}~(b). At this voltage, no visible deterioration of the graphite plates or other indicators of undesired faradaic reactions were observed. 

Apart from the voltage drop, the use of multiple cell pairs will also reduce the specific pumping energy required. In this case, the absolute pumping energy for the flow electrodes will stay the same, while the amount of product stream is multiplied by the number of cell pairs. The contributions of the different resistances, as well as the pumping energy, to the total energy demand is further discussed in Section~\ref{comparison_desalination_pumpingenergy}.

Based on these estimations, the use of multiple cell pairs seems to be a promising approach for FCDI processes, which is in agreement with the observations presented by Ma et al. \cite{Ma2020}) regarding a comparable FCDI stack, which is operable in a semi-continuous manner. In the following, the concept of single-module FCDI with multiple cell pairs is implemented, tested, and its performance evaluated.

\section{Materials and methods} \label{Implementation}

\begin{figure}[h!]
	\centering
    \subfigure{\includegraphics[width=0.8\textwidth]{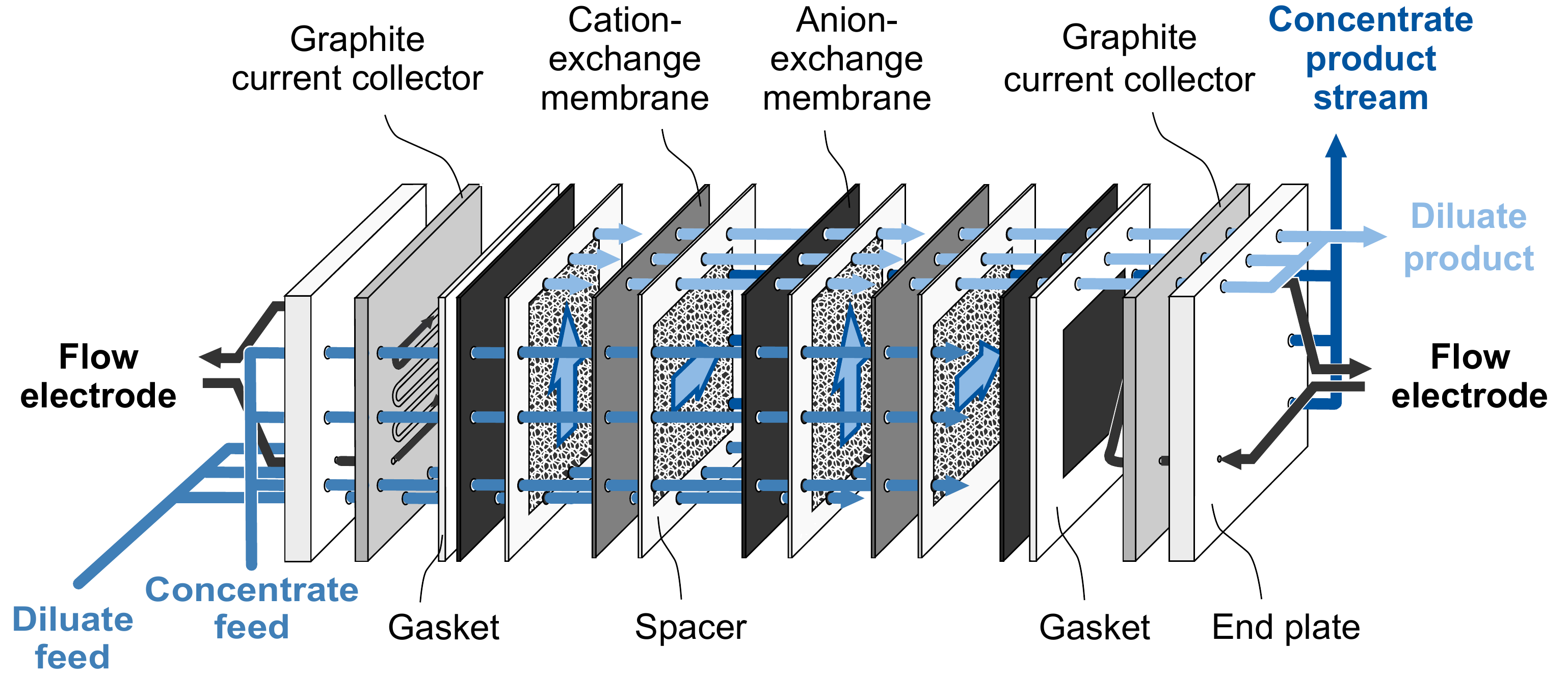}}
\caption{Illustration of the cross-flow module layout applied for multiple cell pair experiments. In this example, a module with two cell pairs is illustrated.}
\label{fig:crossflow_module}
\end{figure} 

\begin{figure}[h]
	\centering
    \subfigure{\includegraphics[width=0.6\textwidth]{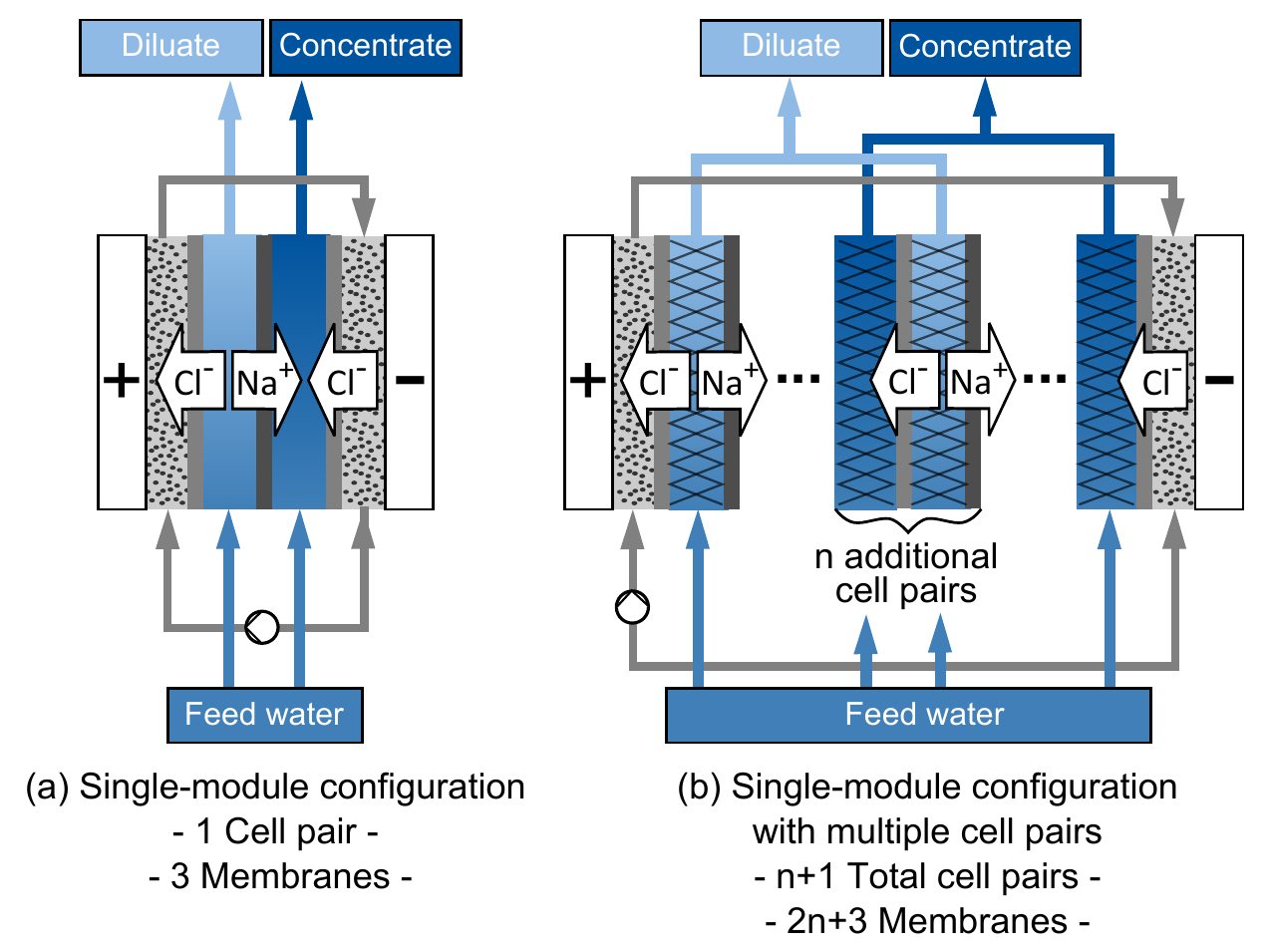}}
\caption{Layout of continuous single module FCDI systems with (a) one cell pair (standard layout), (b) multiple cell pairs (1 cell pair + n additional cell pairs). 
}
\label{fig:Stack_layout}
\end{figure}

The FCDI experiments were performed in cross-flow mode, as illustrated in Figure~\ref{fig:crossflow_module}, which is a simplified 3D representation of the applied cell design. This module design enables the addition of one or more cell pairs into an FCDI module in single-module configuration \cite{Rommerskirchen2015}. Otherwise, the design of the FCDI cells applied in this study is identical to the design described in recent FCDI-related publications by our workgroup \cite{Rommerskirchen2018Energy,Rommerskirchen2019}. The FCDI cells consist of polyethylene end plates, flat gaskets, epoxy-impregnated graphite electrodes (M\"uller \& R\"ossner GmbH \& Co. KG, 180x180x10~mm) with flow channels for the flow electrodes (3~mm width, 2~mm depth, 200~cm overall length), two anion-exchange membranes and one cation-exchange membrane in the configuration with one cell pair
(Fumasep FAB-PK-130/ED-100 and Fumasep FKB-PK-130/ED-100, Fumatech BWT GmbH) with an effective surface area of 121~cm$^2$, and a 0.5~mm mesh spacer (Fumatech BWT GmbH, ED-100). The applied ion-exchange membranes are reinforced with a PEEK-grid and have a thickness of 130~$\mu$m. We applied polyethylene tubing, which is connected to the polyethylene end plates by standard fully polymeric (no metal parts) screw-in tube connectors. The distribution to the three different inlets is implemented via tubing T-connectors outside the cell.
For each added cell pair, two net spacers (ED-100, Fumatech BWT GmbH), one anion-exchange membrane (AEM) (Fumasep FAB-PK-130, ED-100, Fumatech BWT GmbH) and one cation-exchange membrane (CEM) (Fumasep FKB-PK-130, ED-100, Fumatech BWT GmbH) were added between the central CEM and the AEM adjacent to the flow electrode on the concentrate side, as illustrated in Figure~\ref{fig:Stack_layout}. The applied ED-100 net spacers are commercially available spacers with integrated PVC gaskets and a thickness of 500~$\mu$m, which are usually applied inED processes.

The experiments were performed with two Keysight power supplies (Keysight Technologies Inc., E3466A). A digital multimeter (Voltcraft, VC270 Green-Line) was used to verify the actual voltage applied to the graphite current collectors. The flow electrodes were recirculated through the FCDI systems using a peristaltic pump (Ismatec MCP process drive, Masterflex two-channel pump head Easy-Load~II, Norprene tubing). The feed water was supplied by a second peristaltic pump (Ismatec Reglo ICC peristaltic pump with two independent channels). The pumps, power supplies and conductivity sensors (LTC~0,35/23; Sensortechnik Meinsberg) were connected to a custom-made measuring system "ZUMO-FUMS" (ZUMOLab GbR, Wesseling, Germany), which was used to control the pumps and power supplies as well as record the experimental data. Whenever an FCDI module was newly assembled, fixed amounts of water were recirculated separately through each flow electrode, diluate and concentrate compartment for several hours to rule out internal and external leakages.

Experiments were performed with one, two and three cell pairs, as indicated in Table~\ref{tbl:experiments_concepts}. In case of the experiments with several cell pairs, the feed flow rates were multiplied by the number of cell pairs, to achieve comparable hydrodynamic conditions within the cell. Hence, overall flow rates of 1, 2 and 3~mL/min were applied for the diluate feed and 0.04, 0.08 and 0.12~mL/min were applied for the concentrate feed.

\begin{table}[h!]
    \small
	\centering
	\caption{Overview of parameter settings for desalination experiments.}
	\label{tbl:experiments_concepts}
		\begin{tabular*}{5.55in}{cccccc}
			\hline
  &\textbf{Cell} &\textbf{Slurry} &\textbf{} &\textbf{Feed water} &\textbf{}\\
\textbf{Number of}  & Voltage & NaCl & Flow rate & NaCl & Flow rate \\
\textbf{cell pairs}  &  &  &  & & per cell pair \\
\textbf{} & V & g/L & mL/min & g/L & mL/min \\
            \hline 
\textbf{1 CP} & 1.2 &  60  & 150 & 60 & D:1/C:0.04 \\
\textbf{1 CP} & 1.2 \& 1.4 &  120  & 150 & 120 & D:1/C:0.04 \\
\textbf{2 CP} & 1.2 & 60  & 150 & 60 & D:1/C:0.04 \\
\textbf{2 CP} & 1.2\&1.4\&1.8 & 120  & 150 & 120 & D:1/C:0.04 \\
\textbf{3 CP} & 1.2 & 60  & 150 & 60 & D:1/C:0.04 \\
\textbf{3 CP} & 1.4 \& 1.8 & 120  & 150 & 120 & D:1/C:0.04 \\
            \hline 
 		\end{tabular*}
  \vskip 0.1cm	
\end{table}  

The experiments were conducted at constant voltage. In general, the product stream concentration under constant voltage operation can be influenced by the feed concentration and the flow rates of the feed water and the flow electrodes. The addition of further cell pairs leads to an increased overall cell resistance. A moderate desalination rate was chosen in this study to reduce the impact of the cell resistance.  The experiments were performed in such a way that the concentrations in the diluate product stream did not fall below 29.5~g/L. Hence, the cell resistance for additional cell pairs was similar in all experiments with the same NaCl feed concentration, although the diluate concentration varied. In this range (e.g. 29.5~g/L in case of one cell pair and 33.5~g/L in case of three cell pairs), the differences in electrical resistance arising from concentration differences in the diluate stream  have a negligible influence of around 0.002~$\Omega$, which ensures the comparability of the experiments. 

All experiments described in this section were performed with activated carbon slurries used as flow electrodes, which were prepared from 15~wt.\% Carbopal~SC11PG (Donau Carbon GmbH) in a NaCl solution with a concentration equaling the feed concentration, either 60~g/L or 120~g/L (Sodium chloride $\geq$99.8\%, VWR Chemicals). The slurries were stirred overnight before use. For each flow-electrode circuit the slurry was recirculated through the system at a flow rate of 150~mL/min. The slurry storage beakers were continuously stirred using magnetic stirrers. The overall flow rates used for the diluate feed (1~mL/min per cell pair) and the concentrate feed (0.04~mL/min per cell pair) were multiplied by the number of cell pairs to achieve comparable flow conditions within the flow channel. During steady-state operation, several samples of the diluate and concentrate stream were taken and the density measured to confirm the NaCl concentration.

\paragraph{Current-voltage characteristics}
The current-voltage characteristics (IV-curves) of the different systems were investigated by a step-wise change of the applied voltage in a range between 0~V and 2.5~V in steps of 0.1~V. After each change in voltage, the voltage was kept constant for 10~minutes, which corresponds to about two times the residence time. The steady-state electrical current was measured for each step. 

During the measurements, the diluate feed flow rate was kept constant at 1~mL/min. To shorten the time until equilibrium is reached, the concentrate flow rate was set to 1~mL/min. To test the influence of the concentrate feed flow rate on the current-voltage characteristics of the system, one experiment was performed at a concentrate feed flow rate of 0.04~mL/min. The activated carbon content in the flow electrodes was 15~wt.$\%$. 

\paragraph{Experiments comparing FCDI and electrodialysis}
For the experiments one FCDI and one electrodialyses (ED) cell were assembled, which each comprise three cell pairs in the way described above. In this case, the membranes separating the feed streams from the electrolyte rinse solutions or flow electrodes were two cation-exchange membranes. The inner, alternating membranes separating the feed solutions were three anion-exchange membranes and two cation-exchange membranes of the same types as described above. Both cells were based on the ED-100 layout distributed by the Fumatech BWT GmbH. The electrodialysis stack consisted of an ED-100 cell with three cell pairs. Standard ED-100 mesh spacers with a thickness of 0.5~mm (Fumatech BWT GmbH) were used in both cells. The pumps and power sources were the same as described above. 

Two sets of experiments were performed with the above described settings. First, IV-curves were measured using the same procedure as described above. Second, desalination experiments with feed flow rates of 1~mL/min per diluate flow channel and 0.04~mL/min per concentrate flow channel were performed. The FCDI system was operated at a constant voltage of 1.2~V. For the electrodialysis system, the required voltage to achieve the same current as in the FCDI system was determined in a preliminary experiment, following which the desalination experiment was performed at a constant voltage of 3.67~V.
The experimental parameters of the desalination experiments are given in Table~\ref{tbl:experiments_ED}. Samples were taken during steady-state operation of the systems to confirm the salinities of feed, diluate and concentrate solutions.

\section{Assumptions and terminology: Influence of the system boundaries on the energy demand}
\begin{figure}[t]
	\centering
    \subfigure{\includegraphics[width=0.45\textwidth]{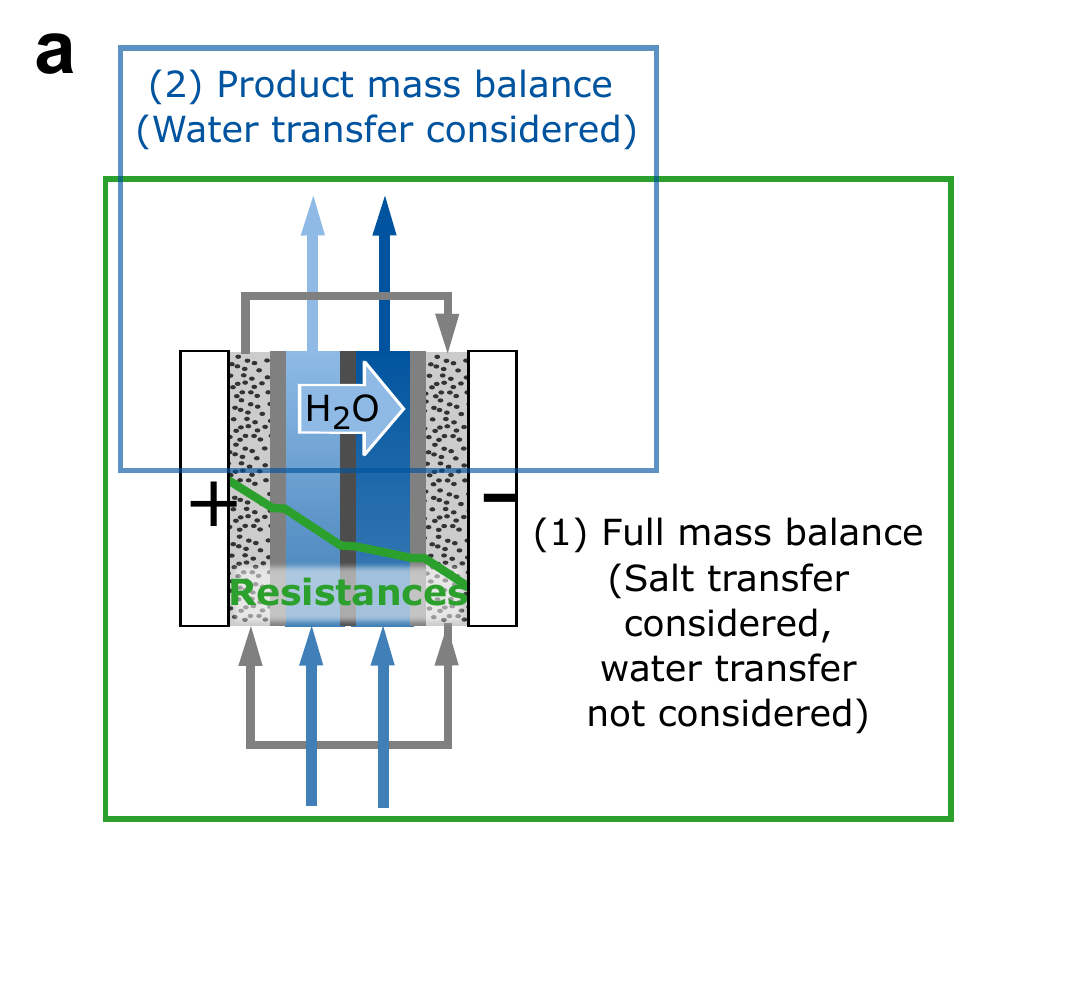}}
    \subfigure{\includegraphics[width=0.45\textwidth]{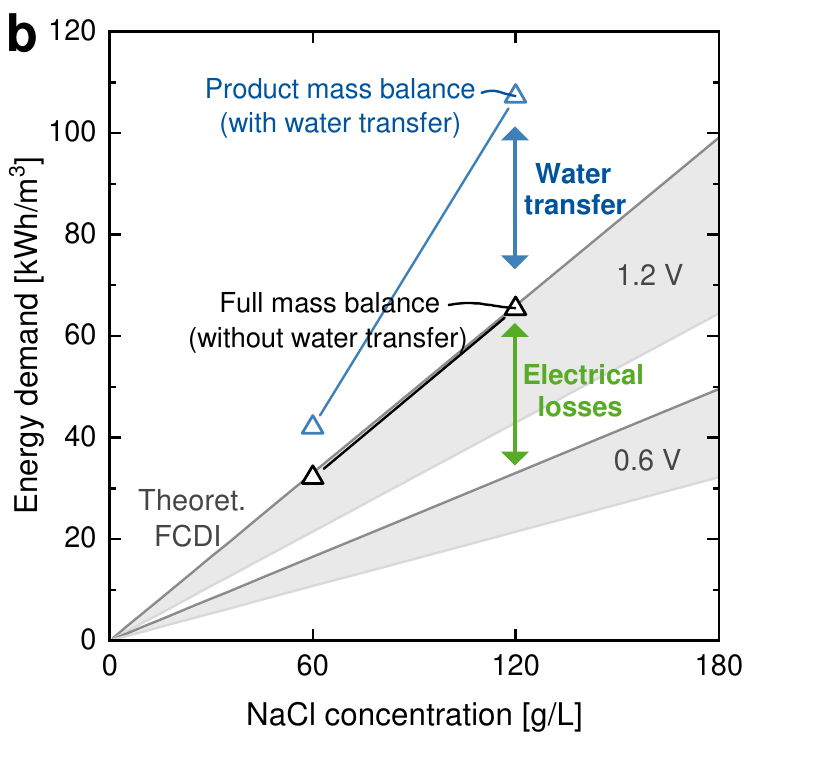}}
\caption{Illustration of (a) the key factors increasing the energy demand in an FCDI process: Water crossover and resistances, which increase the overall cell voltage, and (b) their impact on the electrical energy demand. It depends on the chosen mass balance, as illustrated in (a), whether the water transfer is considered, which leads to a salination of the diluate and a dilution of the concentrate.}
\label{fig:balancing}
\end{figure}

The choice of the system boundary is crucial when analyzing the energy demand of FCDI systems. Two possible approaches are illustrated in Figure~\ref{fig:balancing}~(a). 

In the first approach, the salt transfer is calculated based on the flow rates and salt concentrations of feed and product water streams, indicated by the green system boundary in Figure~\ref{fig:balancing}~(a). In this case, the actual, overall salt transfer is considered. This system boundary is useful for the analysis of the salt transfer and the back diffusion of salt, as expressed in the current efficiency. This means, the energy demand for overcoming the electrical losses/ resistances in the system is considered.
However, due to the additional transfer of water molecules with the ions as well as the osmotic transfer, the dilution of the diluate stream and concentration of the concentration stream is not as high as expected when purely considering the salt transfer. Hence, the system performance is overestimated and the energy demand underestimated in case of the first approach. The energy demand calculated in Wh/g$_{salt}$ is, of course, correct in this case. However, the amount of salt transfered in the system should not be used to calculate the energy demand in kWh/m$^3$ product stream, as in this case the energy demand would be significantly underestimated, as illustrated in Figure~\ref{fig:balancing}~(b), due to the increase of the diluate stream concentration and the dilution of the concentrate stream. \\
In case of the second option, marked by the system boundary in blue in Figure~\ref{fig:balancing}~(a), only the product water flow rates are considered in combination with the feed and product salt concentrations. It is assumed that the volumetric flow rate of the water streams stays constant when passing the cell. This way, the actually applied electrical power, calculated from the applied voltage and the measured current, can be set in relation to the product water streams. Thus, the negative effect of diluate concentration and concentrate dilution due to water crossover is considered in the energy demand, as illustrated in Figure~\ref{fig:balancing}~(b).\\

Due to the influence of the chosen system boundaries, the assumptions and terminology chosen in this article regarding the energy demand and current efficiencies are elucidated in the following.

\subsection{Electrical energy demand per $m^3$ diluate product stream}
There are several ways to calculate the energy demand for the desalination of one cubic meter of salt water, as indicated above. In addition to the differences between product and full mass balance, disparities between the mass balances around the diluate and around the concentrate stream were observed.

\paragraph{Neglect of water transfer, $E_{dil,noWT}$}
When calculating the energy demand based on the full salt mass balance, the water transfer is neglected. The salt stream which is transferred from the diluate stream to the concentrate stream can be calculated in two ways: (1) via the salt mass balance around the diluate stream, which results in Equation~\ref{eqn:EdilDnoWT1} for the energy demand $E_{dil,D,noWT}$, and (2) via the salt mass balance around the concentrate stream, which results in Equation~\ref{eqn:EdilCnoWT2} for the energy demand $E_{dil,C,noWT}$. The average between the two energy demands $E_{dil,D,noWT}$ and $E_{dil,C,noWT}$ is in the following termed $E_{dil,noWT}$ (Equation~\ref{eqn:EdilnoWT3}).

\begin{equation}
  E_{dil,D,noWT} = \frac{c_{Feed} \cdot I \cdot V}{ n_{cp} \cdot (c_{Feed,D} \cdot \dot{V}_{Feed,D}- c_{Diluate} \cdot \dot{V}_{Diluate}) }    \label{eqn:EdilDnoWT1}\\
\end{equation}
\begin{equation}
  E_{dil,C,noWT} = \frac{c_{Feed} \cdot I \cdot V}{ n_{cp} \cdot (c_{Conc} \cdot \dot{V}_{Conc}- c_{Feed,C} \cdot \dot{V}_{Feed,C}) }       \label{eqn:EdilCnoWT2}
\end{equation}

\begin{equation}
  E_{dil,noWT} = \frac{E_{dil,D,noWT} + E_{dil,C,noWT}}{2}       \label{eqn:EdilnoWT3}
\end{equation}

In these equations, $I$ stands for the electric current and $V$ for the cell voltage. $c_{Feed}$ and $\dot{V}_{Feed}$ are the NaCl concentration and volumetric flow rate of the feed stream, in which the subscripts $D$ and $C$ stand for the diluate and concentrate stream, respectively. $n_{cp}$ stands for the number of cell pairs.

The consideration of the full salt mass balance allows the calculation of the actually transferred salt amount. However, the additional transfer of water due to hydration shells and osmotic effects leads to a reduction of the actual desalination and concentration performance. The concentrate is diluted and the diluate stream concentrated due to the transfer of water. This effect is not considered when calculating the energy demand via the full salt mass balance.

\paragraph{Consideration of water transfer, $E_{dil,withWT}$}
When calculating the energy demand based on the product salt mass balance, however, the water transfer is considered. By using only the volumetric flow rate of the product streams, the actual energy demand for the production of a specific volume of product (diluate or concentrate) of a certain desired concentration can be calculated, as given in Equation~\ref{eqn:EdilDwithWT} for the energy demand $E_{dil,D,withWT}$, and in Equation~\ref{eqn:EdilCwithWT} for the energy demand $E_{dil,C,withWT}$. The average between the two energy demands $E_{dil,D,withWT}$ and $E_{dil,C,withWT}$ is in the following termed $E_{dil,withWT}$ (Equation~\ref{eqn:EdilwithWT}).

\begin{equation}
  E_{dil,D,withWT} = \frac{c_{Feed} \cdot I \cdot V}{  (c_{Feed,D} - c_{Diluate} ) \cdot n_{cp} \cdot \dot{V}_{Diluate} }        \label{eqn:EdilDwithWT}
\end{equation}
\begin{equation}
  E_{dil,C,withWT} = \frac{c_{Feed} \cdot I \cdot V}{(c_{Conc} - c_{Feed,C} ) \cdot  n_{cp} \cdot \dot{V}_{Conc}  }       \label{eqn:EdilCwithWT}
\end{equation}

\begin{equation}
  E_{dil,withWT} = \frac{E_{dil,D,withWT} + E_{dil,C,withWT}}{2}       \label{eqn:EdilwithWT}
\end{equation}

By using the above described equations, the additional electrical current required to reach exactly the concentration of 0~g/L NaCl in the diluate product stream is considered. Additional resistances within the desalination cell, especially due to the change of the diluate stream conductivity and the additional resistance of depleted laminar boundary layers at the membrane surface, are not considered by this. This means the values for "full desalination" given in this article are not completely realistic and mostly serve as a base for comparing the different module layouts.

A full desalination is likely not possible using a single cell and single pass FCDI system at feed concentrations of 60~g/L NaCl. A (nearly) full desalination may be achieved when using an optimized cell design, for example by using ion-conductive spacers, similar to continuous electrodeionization systems. However, a desalination to a low concentration of $<$1~g/L is possible. For practical applications, the actual energy demand for a specific separation task should be determined by an experiment in which the desired concentrations are reached. Only this way, or by applying a process model, it can be ensured that interlocking physical and chemical effects are considered. 

\begin{figure}[h]
	\centering
    \subfigure{\includegraphics[width=0.45\textwidth]{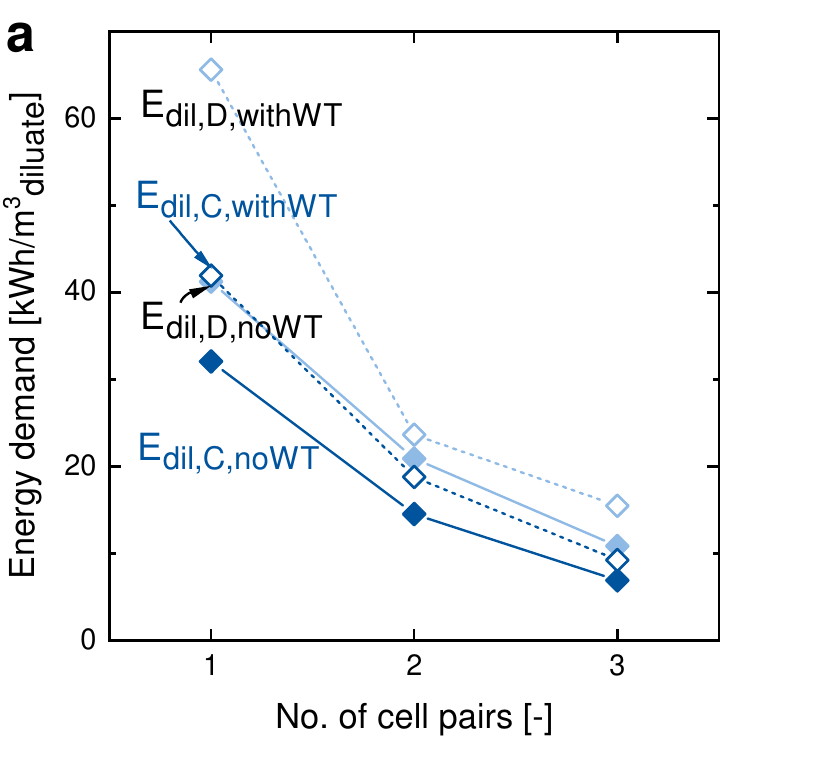}}
    \subfigure{\includegraphics[width=0.45\textwidth]{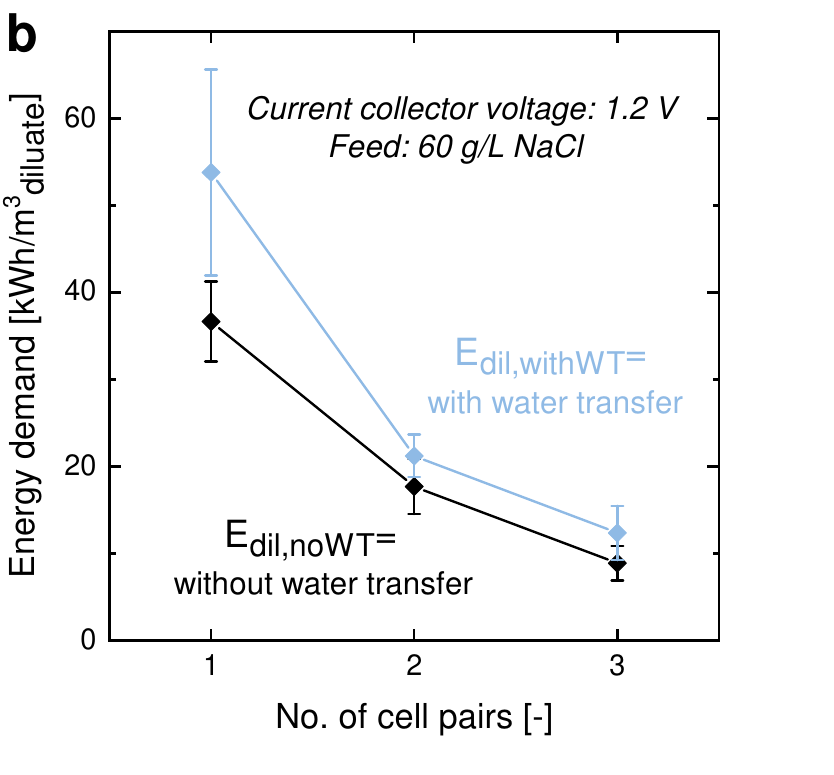}}
\caption{Illustration of the averaging regarding the electrical energy demand calculations in this article. The sample graphs show results of desalination experiments comparing modules with one, two and three cell pairs with a NaCl feed concentration of 60~g/L and a cell voltage of 1.2~V. The graphs show the energy demand calculated per diluate product stream, based on (a) the results for the individual mass balances of diluate and concentrate stream and (b) the average of the two mass balances. The values are recalculated to represent full desalination. Values do not account for pumping energy.}
\label{fig:assumptions}
\end{figure}

\paragraph{Mass balances diluate vs. concentrate} 
Example results for the four different ways to calculate the electrical energy demand are plotted in Figure~\ref{fig:assumptions}~(a). The results for the average electrical energy demand per cubic meter diluate without consideration of the water transfer, $E_{dil,noWT}$, and with consideration of the water transfer, $E_{dil,withWT}$, are plotted in Figure~\ref{fig:assumptions}~(b).

When considering the different curves in Figure~\ref{fig:assumptions}~(a), it is striking that in all cases of this example the energy demand was higher when calculated for the diluate stream, compared to the calculation via the concentrate stream. However, a sighting of data collected over a longer period of time revealed experiments in which also the opposite was the case. The reason for these deviations are most probably experimental and/or measurement errors. An example for a possible systematic error in the data, which leads to a tendency of the energy demand being higher when calculated via the diluate mass balance, may for example be the evaporation of water from the water samples during or after the collection of water samples. This would lead to a concentration of the concentrate (and hence a performance improvement), as well as a concentration of the diluate (and hence a performance reduction and an increased energy demand).

In the following, only the average results between diluate and concentrate mass balance are plotted, similar to Figure~\ref{fig:assumptions}~(b). This averaging based on two mass balances within the same experiment reduces the impact of experimental and measurement errors and allows a more reliable estimation of the energy demand required per cubic meter diluate product stream.

\subsection{Electrical energy demand per $m^3$ concentrate product stream}
The results are more realistic when determining the energy demand based on the product mass balance including the water transfer. Due to this, the energy demand per cubic meter concentrate product stream is only plotted including the water transfer in the following sections. To ensure a comparability between the experimental results, the energy demand was recalculated for a concentration up to 260~g/L NaCl concentration in the concentrate, according to the following equation:

\begin{equation}
  E_{conc260} = \frac{(260~g/L - c_{Feed,C}) \cdot I \cdot V}{(c_{Conc} - c_{Feed,C} ) \cdot  n_{cp} \cdot \dot{V}_{Conc}  }       \label{eqn:Econc260}
\end{equation}

By using this equation, the additional/ reduced current required to reach exactly the concentration of 260~g/L NaCl in the concentrate is considered. When normalizing to 260~g/L all concentrations in this case need to be given in g/L and the other units chosen accordingly. The change in resistances and back diffusion due to changes in the concentration gradients are not considered. To identify the actual energy demand for this specific task, an experiment in which these actual concentrations are reached would be required, due to the many interlocking physical and chemical effects.

\subsection{Current efficiency}
Similar to the energy demand, also the current efficiency (CE) can be calculated based on different mass balances. In case of FCDI processes, the current efficiency brings the transferred salt into relation with the required electric current:

\begin{equation}
  CE = \frac{F \cdot \dot{n}_{Salttransfer}}{ I }    \label{eqn:CE}\\
\end{equation}

In Equation~\ref{eqn:CE}, $F$ stands for the Faraday constant, $\dot{n}_{Salttransfer}$ for the molar amount of salt transferred from the diluate to the concentrate stream, and $I$ stands for the electric current.
The salt flux $\dot{n}_{Salttransfer}$, and hence the current efficiency, can be calculated based on the full mass balance:

\begin{equation}
  CE = \frac{F \cdot (c_{Feed,D} \cdot \dot{V}_{Feed,D}- c_{Diluate} \cdot \dot{V}_{Diluate})}{ I \cdot M_{NaCl}}    \label{eqn:CE2}\\
\end{equation}

Or based on the product mass balance:

\begin{equation}
  CE_{eff} = \frac{F \cdot \dot{V}_{Diluate} \cdot (c_{Feed,D} - c_{Diluate})}{ I \cdot M_{NaCl}}    \label{eqn:CEeff}\\
\end{equation}

In Equations~\ref{eqn:CE2} and~\ref{eqn:CEeff}, $M_{NaCl}$ represents the molar mass of the transferred salt, which in this case is NaCl. The first version of the current efficiency, based on the full mass balance, represents the physically correct current efficiency. The full salt mass balance gives information on how much salt was actually transferred. The CE based on the full mass balance hence represents the actual number of ion pairs transferred per electron and allows conclusions regarding the membrane selectivity. However, as discussed before, this does not represent the apparent current efficiency observed in a real application. In reality, the apparent current efficiency will also be reduced by water transfer, which is why the term "effective current efficiency", $CE_{eff}$, is introduced here for the cases when the CE has been calculated based on the product mass balance. 

\section{Results and discussion}

\subsection{Current-voltage characteristics}
\begin{figure}[h]
	\centering
    \subfigure{\includegraphics[width=0.45\textwidth]{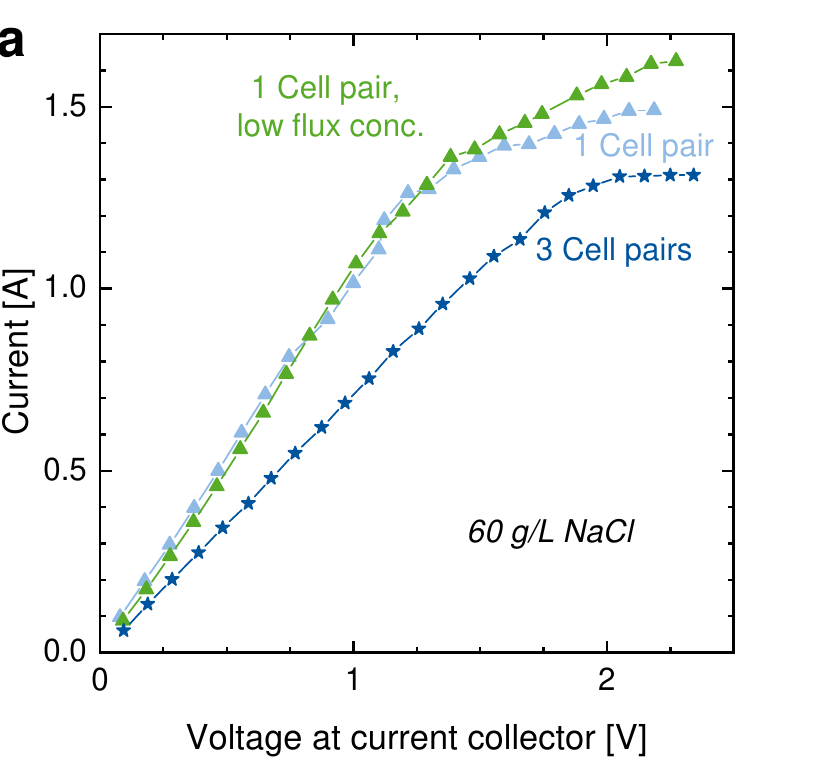}}
    \subfigure{\includegraphics[width=0.45\textwidth]{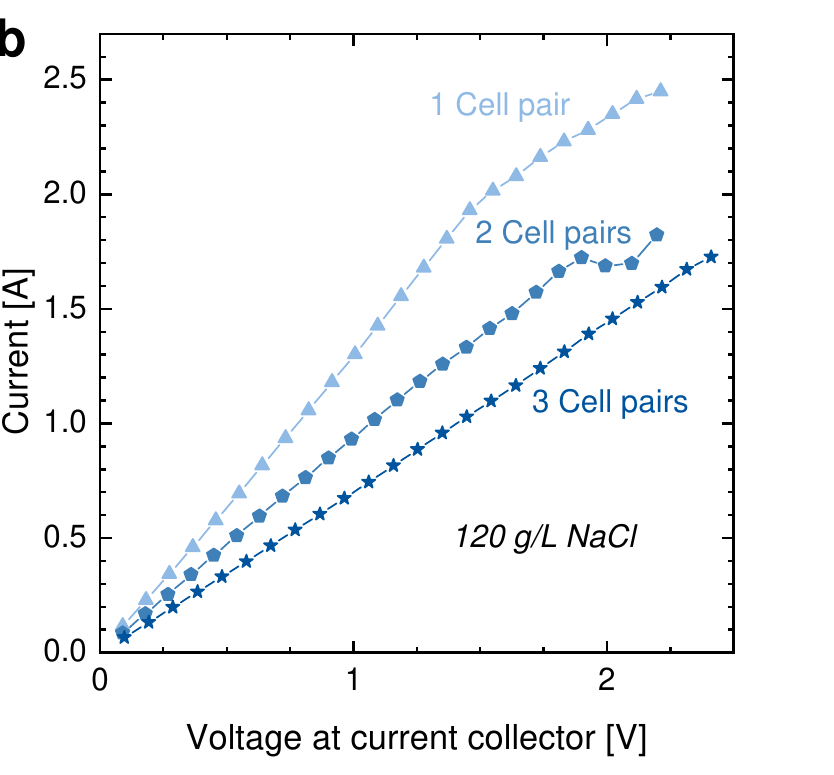}}
\caption{IV-curves of FCDI systems at (a) 60~g/L NaCl, comparing the standard system to a system operated at a low concentrate flux and a system with three cell pairs, and at (b) 120~g/L NaCl, comparing the standard system to systems with two and three cell pairs. The active membrane area is 100~cm$^2$.}
\label{fig:IVcurves}
\end{figure}

For an initial comparison of the performance of different energy demand reduction concepts, the current-voltage characteristics were investigated, as described above. The results for NaCl concentrations of 60 and 120~g/L in feed and flow electrodes are plotted in Figure~\ref{fig:IVcurves} in form of IV-curves.

Similar to IV-curves of electrodialysis systems, the FCDI IV-curves show an ohmic region at low voltages, in which the electrical current increases linearly with increasing potential. At higher potentials, in case of the curves at hand usually around 1-2~V, the curves transcend to a limiting current region. The ohmic resistances and limiting currents differ significantly between the different systems. The resulting ohmic resistance determined from the IV-curves are listed in Table~\ref{tbl:IVcurves_resistances}.

\begin{table}[h]
    \small
	\centering
	\renewcommand{\arraystretch}{0.8}
	\caption{Overview of the ohmic resistances derived from the IV-curves.}
	\label{tbl:IVcurves_resistances}
		\vspace{2mm}
		\begin{tabular*}{5.35in}{lccc}
			\hline
  &\multicolumn{2}{c}{\textit{Experimental parameters}}  &\textit{Analysis results}\\
			\hline  
  &\textbf{Slurry} &\textbf{Feed water}   & \textbf{Ohmic resistances} \\
\textbf{Description}  & wt.\% AC & mL/min & $\Omega$\\
            \hline 
\multicolumn{4}{l}{\textit{IV-curves with 60 g/L NaCl in Feed and Slurry}} \\
            \hline 
\textbf{1 CP, Standard}  & 15   & D:1/C:1                   & 0.927   \\
\textbf{1 CP, low flux concentr.}  & 15   & D:1/C:0.04      & 0.994   \\
\textbf{3 CP}  & 15    & D:1/C:1                            & 1.411   \\
            \hline 
\multicolumn{4}{l}{\textit{IV-curves with 120 g/L NaCl in Feed and Slurry}} \\
            \hline 
\textbf{1 CP, Standard}  & 15   & D:1/C:1                   & 0.789   \\
\textbf{2 CP}  & 15    & D:1/C:1                            & 1.056   \\
\textbf{3 CP}  & 15    & D:1/C:1                            & 1.447   \\
		    \hline
 		\end{tabular*}
  \vskip 0.1cm	
\end{table}  

The standard FCDI system with three ion-exchange membranes (AEM, CEM, AEM) in cross-flow mode has an ohmic resistance of 0.927~$\Omega$. At around 1.2~V, the system starts to transcend to a limiting region, reaching a limiting current ranging around 1.4-1.5~A.  The same system operated at a low concentrate flux, which would enable a high concentration factor in a desalination experiment, exhibits a very similar current-voltage behaviour. The two curves and ohmic resistances are very comparable in the underlimiting region. Only a small difference can be observed in the limiting current region, in which the current of the low concentrate flux curve has a slightly steeper slope, reaching currents of close to 1.6~A in the same voltage range. This insignificant difference may be caused by an increasing salt concentration/conductivity in the concentrate channel and increased back diffusion in case of the reduced concentrate flow rate. 

In case of the standard FCDI system with one cell pair at 120~g/L, the overall behaviour is similar to the system at 60~g/L, while the increased conductivity of the 120~g/L feed solution and flow-electrode electrolyte concentration leads to a significantly lower ohmic resistance (0.789~$\Omega$). Additionally, the limiting current and the voltage at which the limiting region sets in are increased significantly. 

Striking is the unusually flat "limiting current region" in case of many IV-curves plotted here. The reason for this is probably that unlike in IV-curves of electrodialysis processes, the effect causing the "limiting current" region is a different one in case of these FCDI IV-curves. The electric current required for the full desalination of a 60~g/L NaCl solution at a feed flow rate of 1~mL/min, assuming a CE of 1, is 1.65~A. The maximum current achieved in an FCDI IV-curve with a 60~g/L feed is exactly in this range. This indicates that the limiting current observed here is close to the maximum current reachable at the applied flow rate and feed salinity, based on the salt ions available in the feed stream. Having this in mind, the IV-curves can be discussed under a different light.

The introduction of additional cell pairs into an FCDI cell leads, as is to be expected, to an increased ohmic resistance. In case of the 60~g/L NaCl concentration, the resistance increases from 0.927~$\Omega$ (1 cell pair) to 1.411~$\Omega$ (3 cell pairs). This would be equal to around 0.242~$\Omega$ per cell pair. The limiting current decreases slightly with increasing numbers of cell pairs, while the voltage at which the limiting current is reached increases significantly, due to the increased cell resistance. Presuming the reason for the observed "limiting current" is really the maximum desalination, as described above, the cause for the reduced limiting current density in this case may be a maldistribution of the flow between the different cell pairs. In case of small differences in the volumetric flow rate between the different cell pairs the cell pair with the lowest flow rate would reach full desalination at the lowest current density. Hence, the lowest flow rate would limit the further desalination of the other cell pairs. This hypothesis should be investigated further in future works.

The trends at a NaCl concentration of 120~g/L are similar to the trends observed at NaCl concentrations of 60~g/L in the feed solution. In case of the 120 g/L concentration, however, the ohmic resistance increases more significantly with the addition of the second cell pair (which results in a three cell pair system). The reason for this is not clear, a possible explanation may be measurement inaccuracies or an unproportional increase of the resistance when bordered with salt solutions of significantly different concentrations \cite{Geise2013}. 
\subsection{Desalination and concentration performance}
\begin{figure}[h!]
	\centering
    \subfigure{\includegraphics[width=0.45\textwidth]{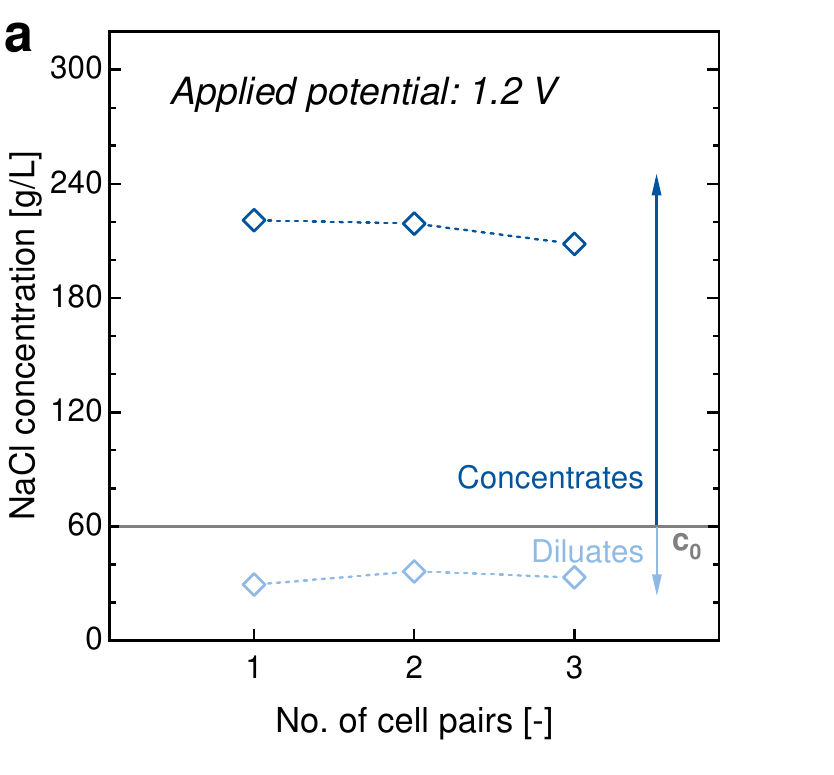}}
    \subfigure{\includegraphics[width=0.45\textwidth]{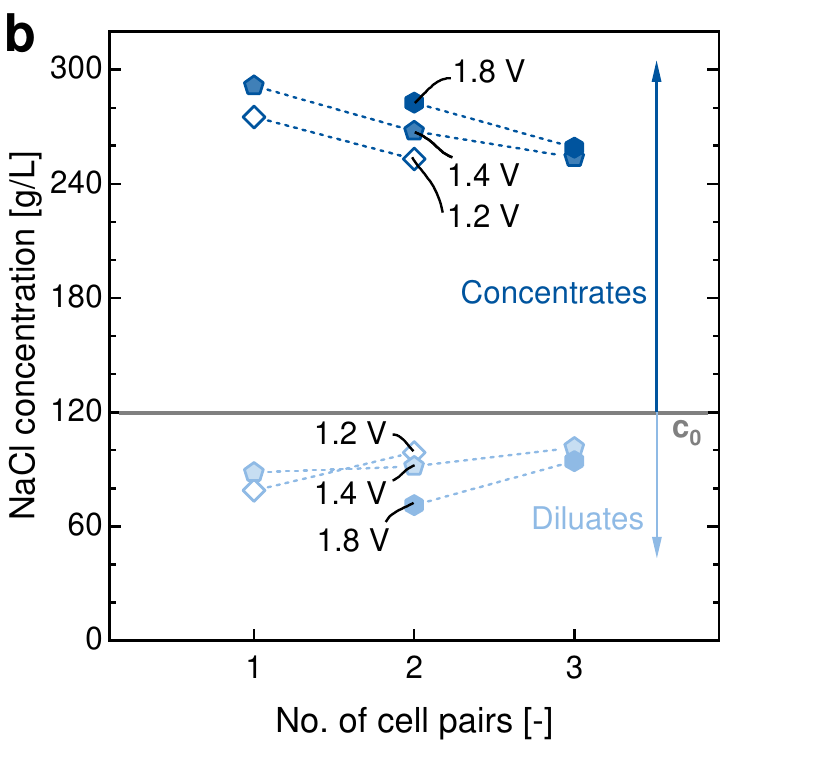}}
\caption{Results of desalination experiments comparing modules with one, two and three cell pairs. The graphs show the NaCl concentrations of feed, diluate and concentrate streams depending on the number of cell pairs at NaCl feed concentrations of (a) 60~g/L and (b) 120~g/L. The potential applied to the current collectors was 1.2~V in case of the 60~g/L experiments and was varied between 1.2~V, 1.4~V and 1.8~V in case of the 120~g/L experiments. }
\label{fig:stack_desalination}
\end{figure}

Figure~\ref{fig:stack_desalination} shows results of desalination experiments performed with the experimental conditions given in Table~\ref{tbl:experiments_concepts}. Figure~\ref{fig:stack_desalination}~(a) shows a plot of the NaCl concentrations of the diluate and concentrate product streams at a feed concentration of 60~g/L and a constant potential of 1.2~V applied to the graphite plates. The grey line indicates the feed concentration c$_{0}$. 
Figure~\ref{fig:stack_desalination}~(b) shows the results of desalination/concentration experiments at a feed concentration of 120~g/L. In this case, the constant potential applied to the graphite plates was varied between 1.2~V, 1.4~V and 1.8~V. 

In case of all experimental results plotted in Figure~\ref{fig:stack_desalination}, the NaCl concentration achieved in the concentrate stream was higher than 200~g/L NaCl. The maximum NaCl concentration achieved in this study was 291.5 g/L. The water recovery in the experiments presented here ranges from 70~\% to 92~\% at NaCl feed concentrations of 60 and 120~g/L NaCl. This is a very promising result regarding the use of FCDI as a potential brine concentration technology.  

In case of both 60 and 120~g/L NaCl as feed concentration, the NaCl concentration was reduced by about 30~g/L when applying the same feed flow rate in the FCDI system. This leads to a diluate product stream concentration of around 30~g/L NaCl in case of the 60~g/L feed stream and about 90~g/L NaCl in the diluate product stream in case of the 120~g/L feed stream at 1.2~V. The NaCl transfer rates in g~NaCl/min calculated for these experiments are in a similar range for both feed concentrations. It is likely that in this concentration range, meaning high salinities of 60-120~g/L NaCl, the additional increase in conductivity has only a small effect. This corresponds well with results presented by Yang et al. \cite{Yang2016FCDIhigh}. The small effect it has is probably countered by a decreasing membrane selectivity at increasing salt concentrations, which results in a decreasing current efficiency and higher back diffusion rates.

In case of both feed concentrations, the desalination and concentration performance decreases when adding additional cell pairs, due to the increased resistance with increasing number of cell pairs. This trend can be avoided to a certain extent by increasing the applied voltage, as shown in Figure~\ref{fig:stack_desalination}~(b). However, as faradaic reactions such as carbon oxidation are to be avoided in FCDI, the applicable voltage is limited. Due to this, the maximum voltage applied in this work did not exceed 2~V in longer experiments and up to 2.5~V in shorter experiments such as the measurement of IV-curves.  

\subsection{Energy demand}

\begin{figure}[h!]
	\centering
    \subfigure{\includegraphics[width=0.45\textwidth]{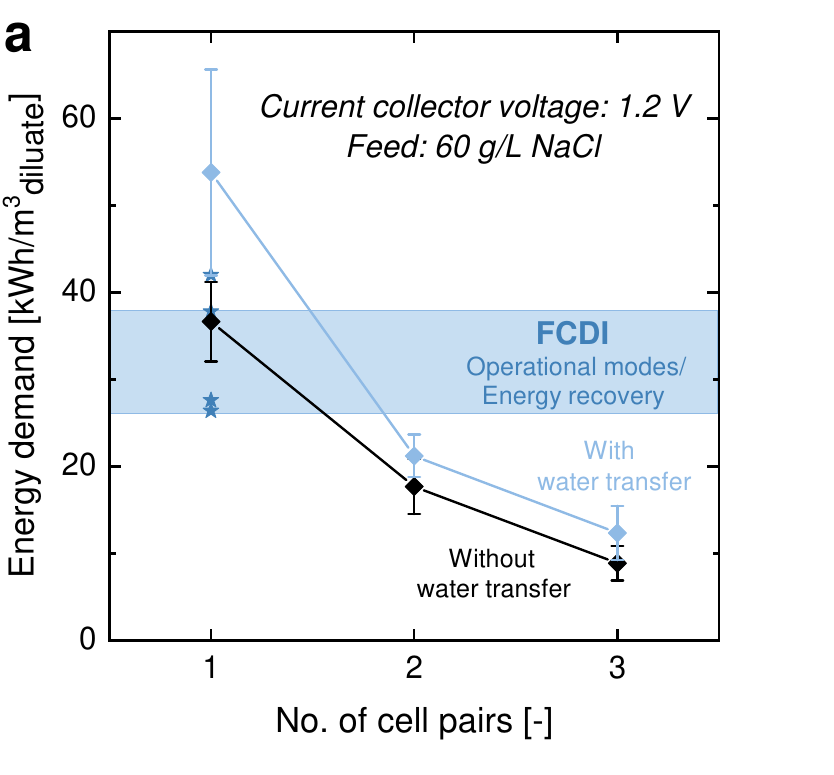}}
    \subfigure{\includegraphics[width=0.45\textwidth]{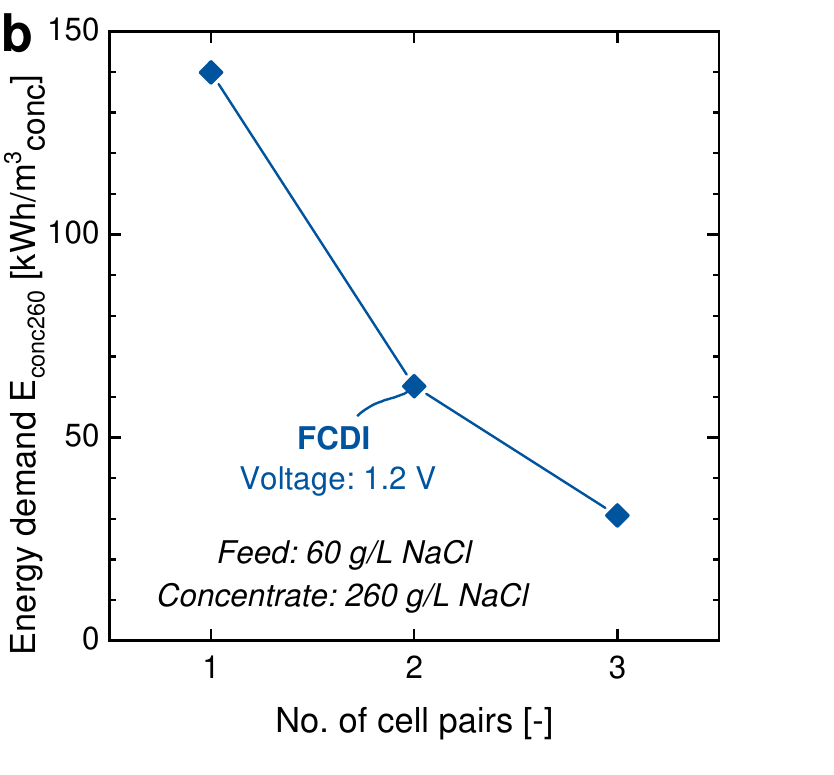}}
\caption{Results of desalination experiments comparing modules with one, two and three cell pairs with a NaCl feed concentration of 60~g/L and a cell voltage of 1.2~V. The graphs show the energy demand calculated (a) per diluate product stream, compared to the value range achieved in our previous article \cite{Rommerskirchen2018Energy}, and (b) per concentrate stream. The values are recalculated to represent full desalination in case of (a) and concentration up to 260~g/L NaCl in case of (b). Values do not account for pumping energy.}
\label{fig:stack_energy_60gL}
\end{figure}

The energy demand corresponding to the results regarding the desalination performance at 60~g/L NaCl in the feedstream, shown in Figure~\ref{fig:stack_desalination}~(a), is plotted in Figure~\ref{fig:stack_energy_60gL}.

The energy demand is normalized per cubic metre diluate product stream, determined according to the equations described above. The energy demand with and without consideration of the water transfer decreases significantly with increasing number of cell pairs. 
When comparing these results to the results presented in our previous article \cite{Rommerskirchen2018Energy}, indicated by the blue area in Figure~\ref{fig:stack_energy_60gL}~(a), it becomes clear that the use of multiple cell pairs can reduce the energy demand of an FCDI process more efficiently than the partial energy recovery achievable using a second regeneration module. With energy recovery, the energy demand for the desalination of a 60~g/L NaCl solution was reduced to around 25~kWh/m$^3$ diluate, while a value of only 15.5~kWh/m$^3$ diluate was reached using an FCDI module with three cell pairs.

The same trend is visible in Figure~\ref{fig:stack_energy_60gL}~(b), which shows the energy demand per cubic metre concentrate product stream. In this figure, all values consider the water transfer, which leads to a dilution of the concentrate. The use of three cell pairs leads to a reduction of the energy demand by more than 70~\%, compared to using only a single cell pair.\\

\begin{figure}[h!]
	\centering
    \subfigure{\includegraphics[width=0.49\textwidth]{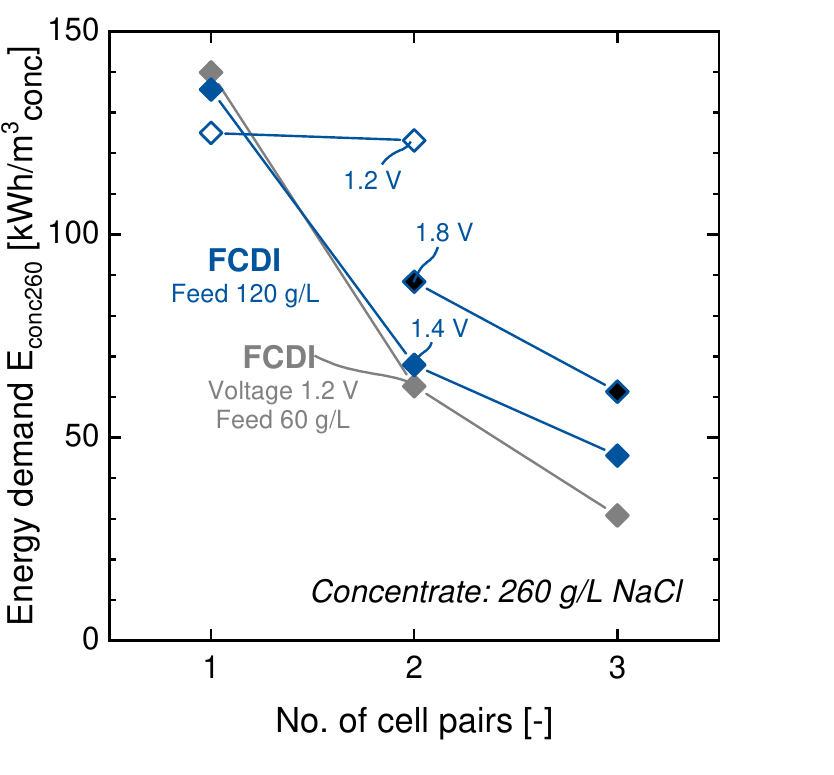}}
\caption{Results of desalination experiments comparing modules with one, two and three cell pairs with a NaCl feed concentration of 120~g/L and a constant cell voltage, which was varied between 1.2~V, 1.4~V and 1.8~V. The graphs show the energy demand calculated per cubic meter concentrate product stream. The values are recalculated to represent concentration up to 260~g/L NaCl. Values do not account for pumping energy.}
\label{fig:stack_energy_120gL}
\end{figure}

Figure~\ref{fig:stack_energy_120gL} shows the energy demand per cubic metre concentrate product stream, which corresponds to the desalination performances plotted in Figure~\ref{fig:stack_desalination}~(b). The results for the concentration of a 120~g/L NaCl solution at varied cell voltages are compared to the results achieved for the concentration of a 60~g/L NaCl solution. All in all, the desalination and concentration of a 120~g/L NaCl solution at a cell voltage of 1.2~V worked well when using a single cell pair. However, when increasing the number of cell pairs to two, the energy demand hardly decreased when applying a voltage of 1.2~V. When applying a voltage of 1.4~V or higher, the energy demand decreases significantly with increasing number of cell pairs, similar to the results observed for a feed concentration of 60~g/L NaCl. A reason for this is possibly a strong back diffusion and increased resistance, which requires the application of a higher voltage. Whether this is a reproducible trend should be investigated further in the future. When the applied voltage is increased further to 1.8~V, the desalination performance improves while at the same time the energy demand increases. However, the overall energy demand is still low compared to a single cell pair system. The right choice of the applied voltage will be an optimization task when designing an FCDI system for an actual application. While increasing the voltage leads to higher current densities, and hence reduced capital cost due to a lower required membrane area, it also leads to a higher electrical energy demand and hence increased operation costs. 

Surprising is that at a constant voltage of 1.4~V, the energy demand for the concentration up to 260~g/L NaCl is nearly the same for a for a 120~g/L feed solution as for a 60~g/L feed solution at 1.2~V. This is a distinct difference to thermally driven processes, in which the feed concentration has a major effect on the energy demand when aiming at the same concentrate concentration. Hence, an advantage of FCDI processes compared to thermally driven processes may be the ability to bridge this wide concentration range with a moderate energy demand.

\subsection{Cell pair resistances and pumping energy} \label{comparison_desalination_pumpingenergy}

\begin{figure}[h!]
	\centering
    \subfigure{\includegraphics[width=0.45\textwidth]{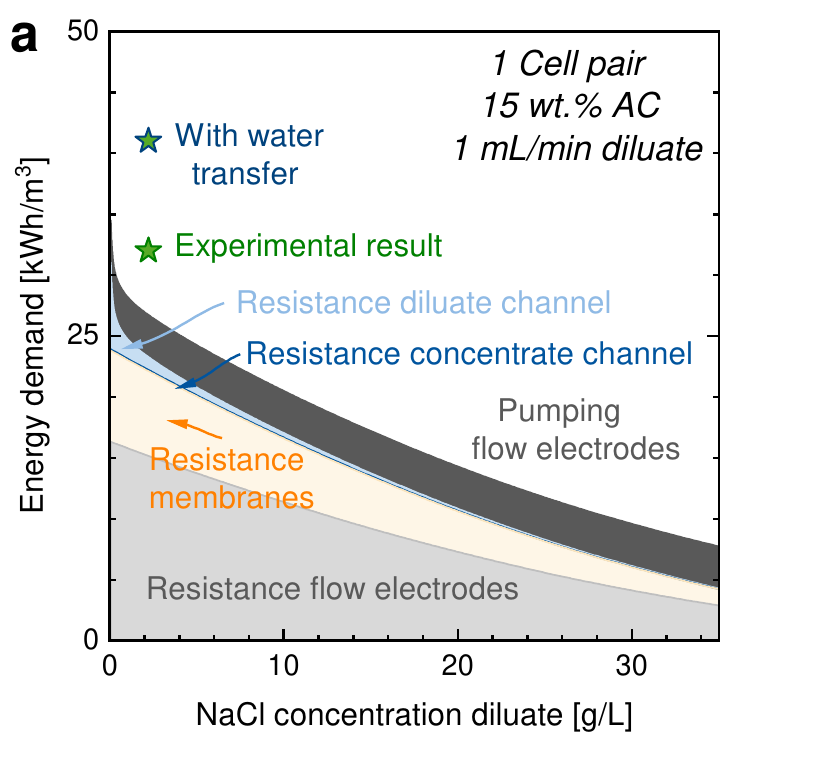}}
    \subfigure{\includegraphics[width=0.45\textwidth]{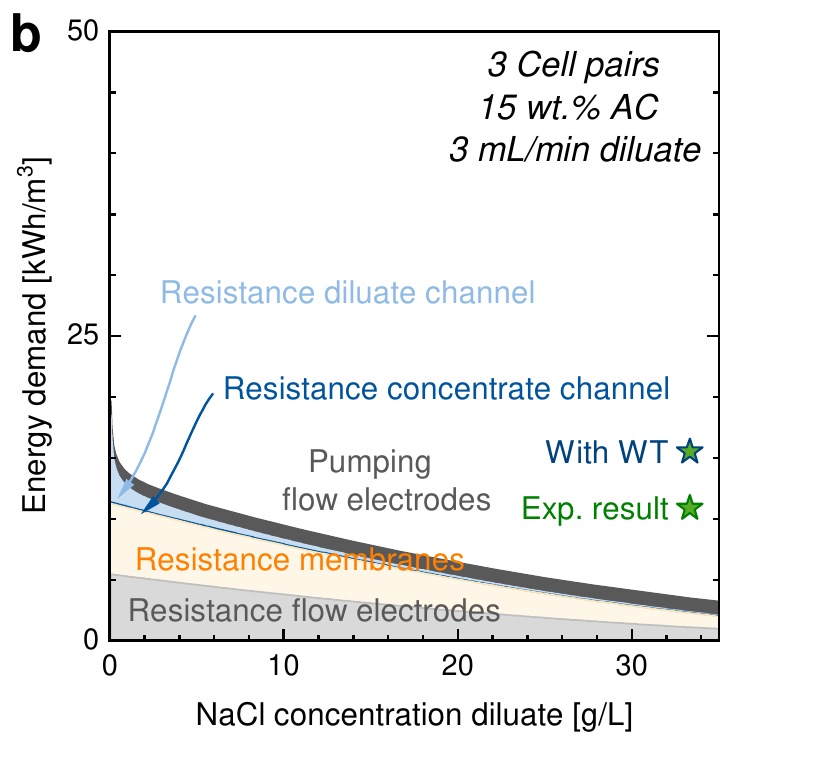}}
\caption{Results of theoretical calculations regarding the influence of the NaCl concentration of the diluate stream on the energy demand and its causes in an FCDI cell with (a) 1 cell pair and (b) 3 cell pairs. Assumptions as described above: Constant diluate flow rate of 1~mL/min, diluate concentration constant throughout cell (overestimated resistances/cell voltage), no consideration of laminar boundary layers, desalination up to the required concentration (varied electrical current). }
\label{fig:FCDI_influences}
\end{figure}

The use of multiple cell pairs in an FCDI module has several potential advantages, as described above. The theoretical calculations presented above in Figure~\ref{fig:concept_stack} illustrate the expected advantages.
The results presented in Figure~\ref{fig:FCDI_influences} are based on the same calculations and illustrate the fractions of the energy demand caused by the different resistances in comparison to the required pumping energy in case of one and three cell pairs. The energy demand calculated for pumping the flow electrodes is based on pressure drop measurements in flow electrodes continuously pumped by a peristaltic pump through a circuit with a dampening vessel to even out pressure fluctuations, and was determined analogue to measurements presented in our previous publication \cite{Rommerskirchen2018Energy}. In Figure~\ref{fig:FCDI_influences}, the theoretical estimations are compared to actual experimental results.

As can be seen from the calculation results plotted in Figure~\ref{fig:FCDI_influences}~(a), the resistance of the flow electrodes is a major driver of the energy demand in case of a single cell pair. Depending on the required diluate concentration, either the membrane resistance or the flow-electrode pumping energy make up a larger part of the total energy demand. By introducing multiple cell pairs, here represented by the example of three cell pairs, the fraction of the flow-electrode resistance and pumping energy, which are two major drivers of the energy demand, can be reduced significantly. This makes the membrane resistance a key factor to reduce the energy demand further in case of an FCDI system with more than one cell pair. Hence, in the following sections the influence of the membranes is investigated in more detail.

When considering the results of practical experiments to the theoretical calculations, the energy demand in the experiments is significantly higher than the estimations. This indicates that not all phenomena are accounted for. Some of these, of course, lie in the nature of the estimation itself: non-ideal behaviour is not considered, such as a reduced current efficiency and back-diffusion. These lead to increased currents in reality. Additionally, laminar boundary layers are not considered, which may increase the required voltage. Studies published by Geise et al. \cite{Geise2013}) and Galama et al. \cite{Galama2014}) also indicate a more than ten times increased membrane resistance at high concentration gradients across the membrane, which may be an explanation for increased cell resistances. The absolute concentration difference across the membrane is in this case at least 170~g/L NaCl. The question is, whether this increased resistance only occurs when one of the bordering concentrations is very low, as described in the study by Geise et al. \cite{Geise2013}, or when there is generally a large concentration gradient. 
The combination of these phenomena may explain why the deviation between experimental and theoretical results is especially significant in case of the three cell pair system. The higher number of membranes in the system may increase the overall effect of the non-idealities which are not considered in this calculation.

\subsection{Effective current efficiency and diluate concentration}

\begin{figure}[h!]
	\centering
    \subfigure{\includegraphics[width=0.49\textwidth]{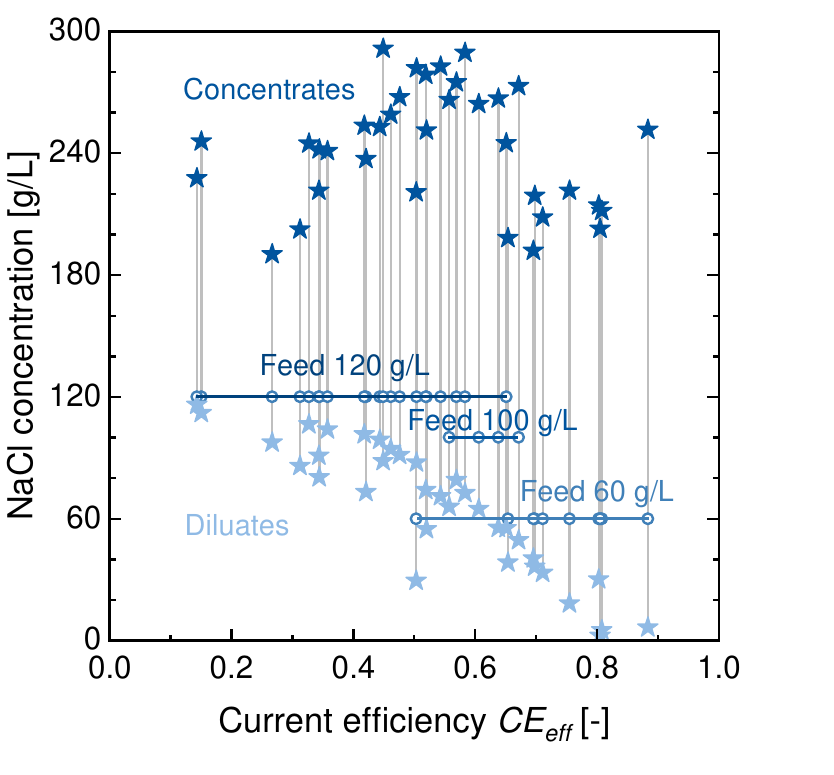}}
\caption{Results for the NaCl concentration in diluate and concentrate product stream plotted over the effective current efficiency, $CE_{eff}$, which considers the water transfer within the system. The plot shows results from a high number of experiments performed with high salinity feeds and various continuous FCDI process and cell configurations.}
\label{fig:CE_allexperiments}
\end{figure}

In some cases, extreme concentration gradients occur over a single ion-exchange membrane in the above described experiments. In literature, increased membrane resistances were observed for high concentration gradients \cite{Geise2013,Galama2014}. Such concentration gradients are rarely reached in electrodialysis processes, due to the choice of higher flow rates and the treatment of lower salinities in most cases. A remaining question is how such high concentration gradients affect the membrane performance and selectivity. Does the current efficiency, for example, mainly depend on the concentration of the diluate stream ("low concentration") or the concentration of the concentrate stream ("high concentration")? The NaCl concentrations measured in diluate and concentrate product streams over the measured current efficiency, obtained from a number of experiments performed in the context of this study, are plotted in Figure~\ref{fig:CE_allexperiments}. The experiments were performed with high salinity feeds and various continuous FCDI process and cell configurations.

There is no clear trend visible for the NaCl concentration in the concentrate streams and the current efficiency. This is surprising, as one may expect a relation due to reduced membrane selectivities at high salinities. However, a very clear trend is visible for the NaCl concentration in the diluate streams and the current efficiency. Low salinities in the diluate stream are clearly associated with high current efficiencies. There are two explanations for this, which may also occur in combination: (1) The low concentrations in the diluate product streams lead to a significant improvement of the membrane selectivity and hence the current efficiency. This would mean the lower concentration bordering an ion-exchange membrane determines the membrane selectivity. This would be a possible advantage for ion-exchange membrane processes aiming at high concentration gradients, in opposition to high diluate flow rates aiming at the reduction of the impact of concentration polarization. (2) Low concentrations in the diluate are only reachable, when also a high membrane selectivity and hence a high current efficiency is achieved. A low membrane selectivity would lead to a stronger back diffusion of salt into the diluate stream and hence a higher salt concentration in the diluate product stream. 

Apart from the scientific interest, the apparent independency of the current efficiency on the concentrate product stream concentration is promising for an application of FCDI processes for brine concentration and brine treatment in general.

\subsection{Energy demand of FCDI compared to mechanical vapor compression}

\begin{figure}[h!]
	\centering
    \subfigure{\includegraphics[width=0.45\textwidth]{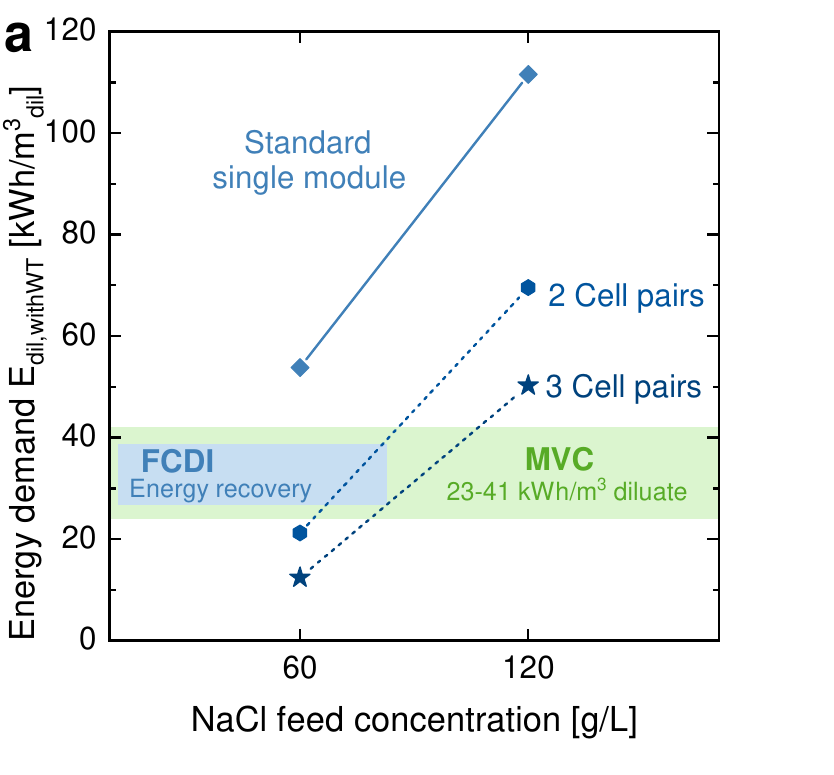}}
    \subfigure{\includegraphics[width=0.45\textwidth]{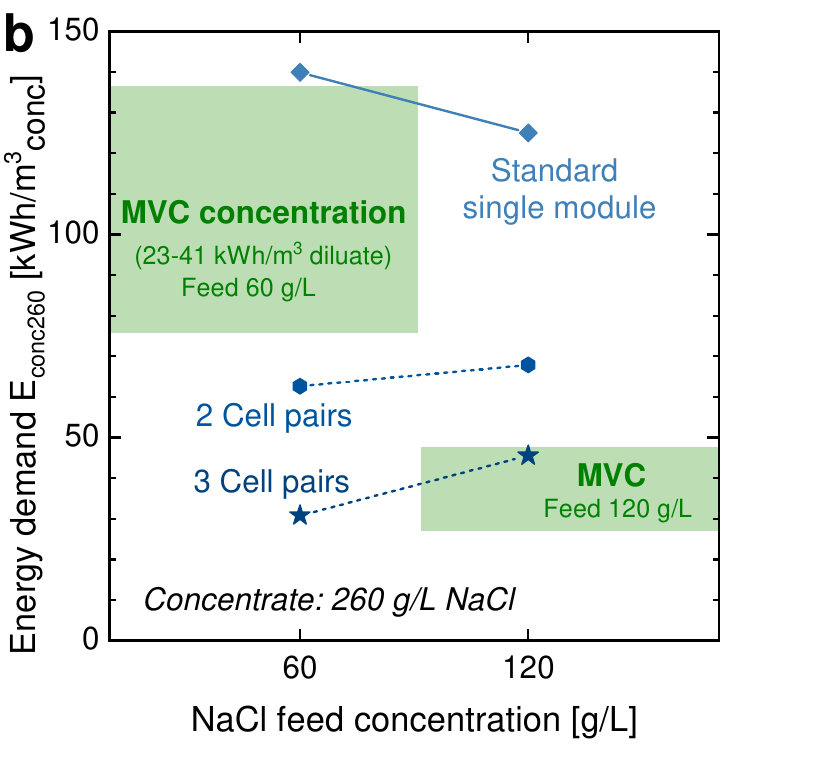}}
\caption{The energy demand of FCDI experiments with one, two and three cell pairs in direct comparison with the energy demand of MVC for a comparable separation task \cite{Thiel2015}) and FCDI with energy recovery \cite{Rommerskirchen2018Energy}.}
\label{fig:FCDI_MVC}
\end{figure}

The improvements achieved by the use of more than one cell pair are significant and illustrate the extant potential of the FCDI technology. Figure~\ref{fig:FCDI_MVC} compares the energy demand achieved in FCDI experiments with one, two and three cell pairs to literature values for mechanical vapor compression \cite{Thiel2015}) and results achieved by FCDI systems with energy recovery presented in our previous article \cite{Rommerskirchen2018Energy}.

The energy demand for the production of one cubic meter of desalinated water using the different FCDI systems compared to mechanical vapor compression (MVC) is plotted in Figure~\ref{fig:FCDI_MVC}~(a). Figure~\ref{fig:FCDI_MVC}~(b) shows results for the energy demand when normalized per one cubic meter of produced concentrate at 260~g/L NaCl. All values consider the increased energy demand due to water transfer from the diluate to the concentrate channel. 

The energy demand scales with increasing feed concentration in case of desalination, which is consistent for all desalination technologies driven by an electrical field. However, in case of FCDI, the energy demand for concentration is fairly independent from the feed concentrations of 60~g/L or 120~g/L NaCl.

To achieve a reduced energy demand it is vital to reduce the cell voltage. This can be achieved by the use of two or more cell pairs, which leads to a division of the voltage drop over the flow electrode by the number of cell pairs. The electrical current stays more or less constant, while the product volume stream is doubled or tripled, leading to a significant reduction of the energy demand per product volume. Compared to FCDI with energy recovery, the use of multiple cell pairs is a more efficient approach, as it enables (1) a better desalination and concentration performance, (2) a higher current density, is (3) generally more stable and reliable in operation, and (4) enables a stronger reduction of the energy demand.

When comparing the lab results regarding FCDI to literature results regarding MVC \cite{Thiel2015}-9, FCDI can compete with MVC regarding the energy demand at the lower feed salinity of 60~g/L. At higher feed salinities, the energy demand for concentration via FCDI is barely able to reach values comparable to MVC. However, considering the given potential of FCDI, this may change in the future. An advantage of FCDI over alternative technologies such as mechanical vapor compression or osmotically assisted reverse osmosis may be the possibility of a selective recovery of ionic components. 

An open question is, whether FCDI can become competitive with state-of-the-art technologies, such as MVC, regarding the investment and operating costs.

\subsection{FCDI compared to electrodialysis}
Electrodialysis (ED) is the most obvious technology, which the continuous FCDI processes presented in this work would be compared to. However, based on the data available in literature, a direct comparison is challenging. Hence, experiments were performed with the aim of collecting directly comparable data for FCDI and ED experiments. The results are described in the following.

\begin{figure}[h!]
	\centering
    \subfigure{\includegraphics[width=0.49\textwidth]{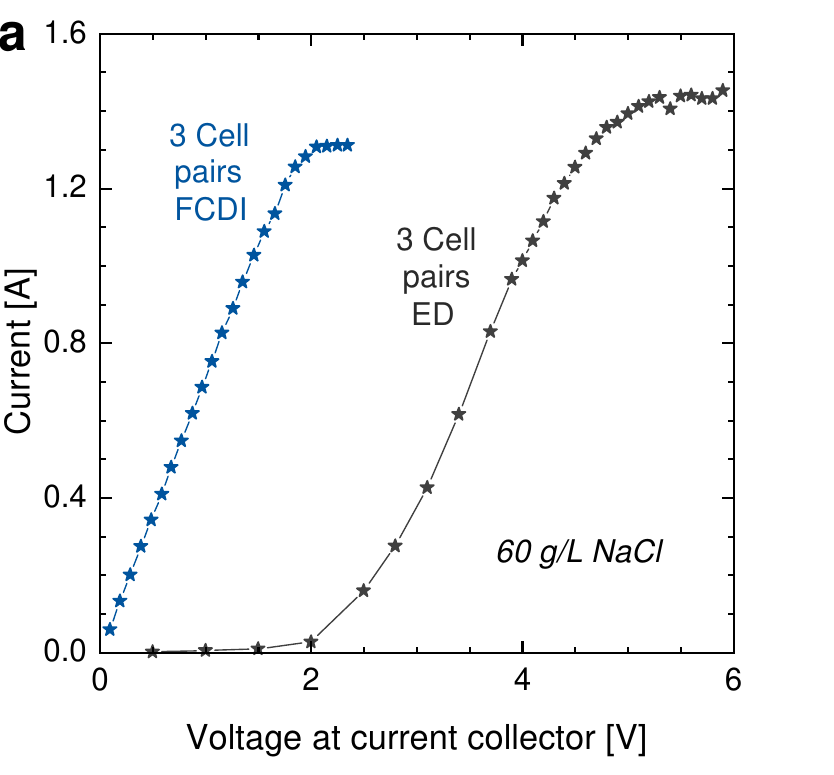}}
    \subfigure{\includegraphics[width=0.49\textwidth]{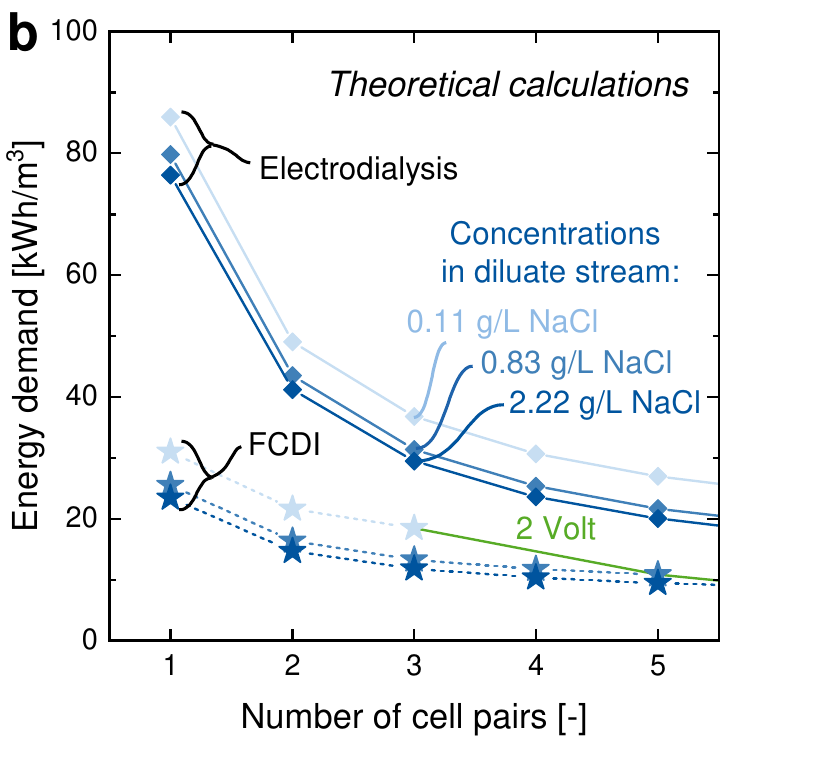}}
\caption{ Direct comparison between an FCDI system and an electrodialysis (ED) system based on the same membrane type, spacer type, cell pair number and feed flow rates. (a) Experimentally determined current-voltage characteristics in comparison and (b) results of theoretical calculations comparing an FCDI to an ED system based on the same assumptions applied for Figure~1. The green line (2 Volt) indicates an estimation of the maximum number of cell pairs viable for reaching the indicated diluate product streams concentrations when assuming 2~V as the maximum voltage applied to the electrodes in an FCDI system. }
\label{fig:FCDI_ED}
\end{figure}

Figure~\ref{fig:FCDI_ED} shows experimental results and theoretical calculations directly comparing an FCDI and an ED system. Figure~\ref{fig:FCDI_ED}~(a) shows the resulting IV-curves of the elctrodialysis and the FCDI system. The results of theoretical calculations presented in Figure~\ref{fig:FCDI_ED}~(b) are based on the same assumptions as described above for Figure~\ref{fig:FCDI_influences_1}.

The IV-curves plotted in Figure~\ref{fig:FCDI_ED}~(a) illustrate the main difference between FCDI and ED well. The slopes in the ohmic region are very similar and the systems reach a very similar limiting current density. The difference in the limiting current density likely results from a small deviation in the feed water flow rate setting in the ED system, which is designed for much higher flow rates. However, while the FCDI system produces a linear IV-curve which goes through the origin in the Ohmic region at lower voltages, there is basically no electrical current flowing through the electrodialysis system at low voltages. At around 2~V, the electrical current starts to increase, passing over into a linear increase in the ohmic region. The reason for this is the overpotential required by the electrodes of the electrodialysis system. In contrast to the FCDI system, which relies on the capacitive adsorption of ions on the flow-electrode surface, the electrodes of the electrodialysis system require faradaic reactions to achieve the transition from ionic current to electrical current. \\

\begin{table}[h]
    \small
	\centering
	\caption{Settings and results of desalination experiments with an FCDI and an electrodialysis setup in direct comparison.}
	\label{tbl:experiments_ED}
		\begin{tabular*}{5.85in}{lccc}
			\hline
  &\textbf{} &\textbf{FCDI} &\textbf{Electrodialysis}\\
            \hline 
Electrical current & \textit{A}  & 0.87 & 0.87  \\
\textbf{Potential at current collectors} & \textit{V} & 1.2 & \textbf{3.67}  \\
Feed concentration & \textit{g/L NaCl} &  60  & 60 \\
Diluate concentration & \textit{g/L NaCl} &  33.3  & 35.6 \\
Concentrate concentration & \textit{g/L NaCl} &  208.3  & 204.4 \\
Number of cell pairs & \textit{-} &  3  & 3 \\
Diluate feed flow rate (per flow channel) & \textit{mL/min} &  1  & 1 \\
Diluate product flow rate & \textit{mL/min} &  0.87  & 0.82 \\
Concentrate feed flow rate (per flow channel) & \textit{mL/min} &  0.04  & 0.04 \\
Concentrate product flow rate & \textit{g/L NaCl} &  0.25  & 0.18 \\
Current efficiency & \textit{-}  & 0.98 & 0.97\\
\textbf{Energy demand desalination} & \textit{kWh/m$^3$} & 14.93 & \textbf{53.01}  \\
\textbf{(incl. water transfer)} & \textit{} &  & \textbf{   }  \\
		    \hline
 		\end{tabular*}
  \vskip 0.1cm	
\end{table} 

Table~\ref{tbl:experiments_ED} shows the results of the desalination experiments directly comparing FCDI and ED. The results show that the FCDI and ED system were able to achieve nearly identical desalination and concentration results and current efficiencies when operated at the same electrical current density. The main difference is, as seen before in the IV-curves, the significantly higher voltage required to run the faradaic reactions at the electrodes. For this specific experiment, the ED system required a voltage increase of nearly 2.5~V in comparison to the FCDI system. This leads to a significant increase in the energy demand.\\ 

To reduce the impact of the electrode voltage drop, the electrodialysis stacks used for industrial applications usually consist of dozens or even hundreds of cell pairs. This way, the voltage drop over the electrode does not have such a significant impact on the overall energy demand anymore. However, such large stacks suffer from different restrictions, such as ionic shortcut currents through feed and drain channels \cite{Veerman2008}. Due to this, it may be advantageous in some cases to be able to operate a desalination/concentration cell efficiently and economically even when using a low number of cell pairs, using a well-designed cell. The plots in Figure~\ref{fig:FCDI_ED}~(b), which show the results of theoretical calculations of the electrical energy demand for FCDI and ED systems depending on the number of cell pairs, illustrate this. For both systems, a significant reduction of the energy demand is achievable by increasing the number of cell pairs to more than one. In case of FCDI, the voltage drop over the electrodes is much smaller and does not require a starting potential, which leads to the significantly lower energy demand and the lower decrease with increasing number of cell pairs. As per the theoretical calculations, the increase to more than three cell pairs improves the energy demand only to a small extent. In case of electrodialysis, a much higher number of cell pairs is theoretically required to reach the same energy demand as FCDI. At a number of five cell pairs, the energy demand of the electrodialysis system is similar to the FCDI system at one cell pair, not accounting for the pumping energy.  \\

Apart from this, electrodialysis stacks require electrochemical reactions at the electrodes to convert the electrical current in the outside circuits into the ionic current within the cell. Next to gas evolution and the associated safety risks, the chemical reactions occuring at the electrodes can lead to unwanted side reactions, which may be undesired in case of specific applications, e.g. involving susceptible chemicals or mixtures. In this regard, FCDI processes may be advantageous for certain applications.

\section{Conclusions}
In this article, it is shown that continuously operated FCDI systems can be applied for the concentration of salt brines. Concentrations of up to 291.5~g/L NaCl were reached in the concentrate product stream at water recoveries ranging from 70~\% to 92~\% at NaCl feed concentrations of 60 and 120~g/L NaCl. Hence, FCDI is a promising technology for the treatment and recovery of salts from high salinity water streams (brines).

So far, the energy demand of FCDI systems tested for brine treatment was comparably high. However, the use of multiple cell pairs in FCDI systems, as presented in this article, is a very promising approach to reduce the overall energy demand of FCDI systems. Advantages of this approach are: (1) Reduced influence of the electrical resistance of the flow electrodes and hence a significantly decreased electrical energy demand, (2) reduced flow-electrode pumping energy per volume product stream and per mass of transferred salt, (3) reduced investment costs compared to stacking of single-cell pair systems. 

The economic feasibility of FCDI systems still needs to be proven. However, with further application-oriented research, FCDI systems may soon become viable for many industrial and agricultural applications aiming at closing material cycles via salt recovery from waste water streams.


 \section*{Acknowledgements}
This work was supported by the German Federal Ministry of Education and Research (BMBF) under the project “ElektroWirbel” (FKZ 13XP5008) and by the European Research Council (ERC) under the European Unions Horizon 2020 research and innovation program (694946). The authors thank Fumatech BWT GmbH and Donau Carbon GmbH for the provision of material samples.
The authors thank the ElektroWirbel project partners and all other project partners and colleagues for the valuable discussions. The authors are especially grateful for the in-depth discussions with Yuliya Schie\ss er, Matthias Woyciechowski, Tomas Klicpera and Youri Gendel.
The authors thank Max D\"uppenbecker, Jingyu Xie, Zhaowei Zou and Niklas K\"oller for their constructive effort in this study. 

\bibliographystyle{elsarticle-num}
\bibliography{lit_FCDIStack_Paper}

\end{document}